\documentclass[ aps, showpacs, showkeys, nofootinbib, floatfix, superscriptaddress]{revtex4}

\usepackage{amsfonts}

\usepackage{amssymb}
\usepackage{amsmath}
\usepackage{graphicx}
\usepackage{epstopdf}

\begin{document}
\title{Study on high-frequency quasi-periodic oscillations in rotating black bounce spacetime}
 \author{Shining Yang}
 \email{yangshining1996@163.com}
 \affiliation{Department of Physics, Liaoning Normal University, Dalian 116029, P. R. China}
 \author{Jianbo Lu}
 \email{lvjianbo819@163.com}
 \affiliation{Department of Physics, Liaoning Normal University, Dalian 116029, P. R. China}
 \author{Xinping Yu}
 \affiliation{Department of Physics, Liaoning Normal University, Dalian 116029, P. R. China}
 \author{Jingyang Xu}
 \affiliation{Department of Physics, Liaoning Normal University, Dalian 116029, P. R. China}

\begin{abstract}
  This study investigates the dynamical effects of particles orbiting a celestial body in rotating Simpson-Visser (RSV) spacetime. The results show that, compared to Kerr and rotating regular black holes, the innermost stable circular orbit (ISCO) of an RSV wormhole is closer to the central object. Using high-frequency quasi-periodic oscillation (HFQPO) data from microquasars and $\chi^2$ analysis, we constrain the spin of microquasars and the regularization parameter $l$ in RSV spacetime based on four HFQPO models and their variants, and evaluate the models using the Akaike Information Criterion and Bayes factor.
  Calculations indicate that $\mathrm{ER}_0$, $\mathrm{ER}_1$, $\mathrm{ER}_2$, $\mathrm{RP}_0$, $\mathrm{RP}_2$,  and $\mathrm{WD}$ models have the same support by observational data as the best model $\mathrm{TD}$. Other models considered in this paper have less or no support from observational data. Concretely, for cases of $\mathrm{ER}_0$ and $\mathrm{RP}_0$ models, the observational constraints on RSV regularization parameter are respectively: $l^* = 0.908_{-0.073}^{+0.086}$ and $l^* <0.314$ at $68 \%$ confidence level, which correspond to the regular or the Kerr ($l^*=0$) BH. For $\mathrm{ER}_1$, $\mathrm{ER}_2$, $\mathrm{RP}_2$, $\mathrm{TD}$, and $\mathrm{WD}$ models, the observational data suggest that RSV objects should be the traversable wormhole, e.g. we have the limits: $l^* =1.850 \pm 0.036$, $l^* =4.964 \pm 0.046$, etc.

\end{abstract}

\keywords{ Black Bounce; quasi-periodic oscillations; wormholes.}

\maketitle

\section{$\text{Introduction}$}
In recent years, the research into physics of black holes (BH) has made significant progress. Observations of gravitational waves \cite{1} and shadow images of black holes \cite{2,3} have confirmed the predictions of general relativity (GR). However, the existence of black holes predicted by GR inevitably leads to singularity in spacetime, causing the classical laws of physics to fail. Although it is generally believed that quantum gravity can solve this problem, the theory of quantum gravity is not yet perfect. So, phenomenological models with singularity-free are widely welcomed. In this context, researchers have proposed many phenomenological models, such as singularity-free gravitational collapse \cite{4,5,6,7,8}, singularity-free cosmology \cite{9,10,11,12,13,14,15}, and regular black holes \cite{9,10,11,12,13,14,15,16,17,18,19,20,21,22,23,24,25}.

The regular BH model was first proposed by Bardeen in 1968. Building on this idea, Simpson and Visser proposed a spacetime metric known as the "black bounce" \cite{25}. This spacetime can unify the description of the Schwarzschild solution, regular black holes, and traversable wormholes \cite{26}. Recently, Mazza, Franzin, and Liberati have extended the Simpson-Visser (SV) metric to a rotating form \cite{27}. The rotating Simpson-Visser (RSV) spacetime is a modified form of Kerr spacetime, which can represent rotating regular BH or rotating traversable wormholes and reduces to the Kerr black hole when the parameter $l=0$. The black-bounce family has received widespread attention due to its unique spacetime structure. Many studies have been conducted on SV spacetime and its rotating forms, such as quasi-periodic oscillations (QPOs), gravitational lensing effects, quasi-normal modes, shadows, and accretion disks \cite{26,27,28,29,30,31,32,33,34,35,36,37,38,39,40,41,42,ysn}.

Besides BH, wormhole is another important theoretical prediction of general relativity. Although the construction of traversable wormholes in GR violates the null-energy condition \cite{43,44,45}, this problem can be solved by quantum gravity or considering the modified gravity theory of GR \cite{ec1,ec2}. Therefore, exploring the observable signals from wormholes has become an important aspect of research on wormhole physics in recent years \cite{46,47,48}. Although there are no unambiguous observations to date that prove the existence of wormholes, observable features such as gravitational lensing \cite{49,50,51,52}, shadows \cite{53,54,55,56}, ironline profile \cite{57}, Lense-Thirring precession \cite{58,59}, and accretion disk radiation \cite{60,61} can be obtained in the wormhole solutions of GR and its modified gravity theory. Research has shown that in some cases wormholes can perfectly reproduce the observational characteristics of Kerr black holes, thus wormholes are also known as black hole imitators \cite{62,63}. This makes it difficult to distinguish between wormholes and black holes in certain situations. Therefore, testing strong gravitational fields through astronomical observational data, studying the deviation of wormholes relative to Kerr black holes, and the different effects induced by this deviation can provide a theoretical basis for identifying different types of celestial bodies \cite{49,64,65,66,67,68}.

Quasi-periodic oscillations have become one of the most powerful tools for testing gravitational theory due to their high-precision characteristics \cite{69,70,71,72,73,74,75,76}. QPOs correspond to peaks observed in radio to X-ray bands of the electrical spectrum of compact objects \cite{77}, which can be divided into low-frequency and high-frequency QPOs based on the observed oscillation frequency. If the observational data can be correctly interpreted with some theoretical models, QPOs can provide evidence for the existence of more compact objects \cite{59}. Many theoretical models have been proposed to explain the observed quasi-periodic oscillation phenomenon. However, so far, no model has been able to fit the observational data from different sources simultaneously \cite{76,77,78,79}. As a result, the exact physical mechanism of generating HFQPO (high-frequency quasi-periodic oscillations) are generated remains elusive. In this article, we investigate the difference in some aspects among the spacetimes of RSV regular black holes, traversable wormholes, and Kerr black holes. We explore these issues by combining observational data of microquasars with considering the popular models: epicyclic resonance (ER) \cite{80,81}, relativistic precession (RP) \cite{82,83}, tidal disruption (TD) \cite{84,85,86}, and warped disc (WD) \cite{87}. At the same time we analyze the possible physical mechanisms for the generation of HFQPO around three types of axially symmetric SV celestial bodies.

The structure of the paper is as follows. The first part is the introduction. The second part of the article provides a brief introduction to the RSV line element, which can uniformly describe some typical black holes and traversable wormholes. In the third and fourth parts, we study the existence range and stability of circular orbits of particles moving on the equatorial plane for Kerr black holes, regular black holes, and traversable wormholes in RSV spacetime. We also derive the general expressions for the radial and latitudinal angular frequencies of particles oscillating in this orbit and provide the position of the innermost stable circular orbit (ISCO). In the fifth section, we demonstrate the possible values for the frequencies of the twin peaks and the locations where the 3:2 resonance occurs for particles oscillating on stable circular orbits under the different four popular HFQPO models (ER, RP, TD, WD) and their variants. In the sixth section, under the aforementioned HFQPO models, we apply the HFQPO data observed from three microquasars (GRS 1915+105, XTE 1550-564, and GRO 1655-40) to constrain the regularization parameter and the spin parameters of the three microquasars in RSV spacetime. Considering that the related result largely depends on the chosen models, it is significant to evaluate various HFQPO models through statistical analysis of observational data. Therefore, we apply the Akaike Information Criterion and Bayes factor to select models with the help of the observational data. The seventh part is the conclusion of the article.

\section{$\text{The rotating Black Bounce metric}$}

 We consider the rotating form of the SV metric, also called the rotating Black Bounce spacetime \cite{27}
\begin{equation}
ds^2=-\left(1-\frac{2 M \sqrt{r^2+l^2}}{\Sigma}\right) dt^2+\frac{\Sigma}{\Delta} dr^2+\Sigma d \theta^2-\frac{4 M \alpha \sin ^2 \theta \sqrt{r^2+l^2}}{\Sigma} dtd \phi+\frac{Asin^2 \theta}{\Sigma} d \phi^2, \label{1}
\end{equation}

with
\begin{equation}
\begin{gathered}
\Sigma=r^2+l^2+\alpha^2 \cos ^2 \theta, \\
\Delta=r^2+l^2+\alpha^2-2 M \sqrt{r^2+l^2}, \\
~A=\left(r^2+l^2+\alpha^2\right)^2-\Delta \alpha^2 \sin ^2 \theta.
\end{gathered} \label{2}
\end{equation}
Here $l$, $M$, $\alpha$ are the regularization parameter, mass, and spin parameter of the central object, respectively. The domain of the radial coordinate $r$ is $r \in(-\infty,+\infty)$. When $\alpha=0$ or $l=0$, the aforementioned line element is reduced to the SV metric or the Kerr metric.
It can be noted that the $r=0$ surface is an oblate spheroid with a size equal to $l$ (BoyerLindquist radius). When $l=0$, the spheroid collapses into a ring at $\theta=\pi / 2$, recovering the Kerr ring singularity. When $l \neq 0$, the singularity is removed, and $r=0$ becomes a regular surface of finite size that observers can traverse. In this case, the metric (\ref{1}) describes a wormhole with its throat located at $r=0$ \cite{27}.
The event horizon is located at $\Delta=0$
\begin{equation}
r_{ \pm}=\left[\left(M \pm \sqrt{M^2-\alpha^2}\right)^2-l^2\right]^{1 / 2}. \label{3}
\end{equation}
Evidently, in the RSV spacetime, when $l=M\pm\sqrt{M^{2}-\alpha^{2}}$, the horizon and the throat coincide, both located at $r=0$, which is also the position of the Kerr ring singularity. With the influence of parameters, the RSV spacetime can describe various types of celestial bodies \cite{9,27}, including traversable wormholes, a one-way wormhole with a null throat, single horizon regular black hole, double horizon regular black hole, a regular black hole with one horizon and a null throat, an extremal regular black hole.

In this article, to compare with the Kerr black hole, we constrain the range of spin parameter:
 $\alpha \in[-M,M]$, and  focus on the range of values of the parameters $\alpha$, $l$, $M$ when describing Kerr black holes, regular black holes, and traversable wormholes. Through the formula (\ref{3}), it can be seen that when  $\l^* \leq \sqrt{-{\alpha^*}^2+2+2 \sqrt{-{\alpha^*}^2+1}}$, the central object described by the formula (\ref{1}) is the Kerr black hole or a regular black hole; when $\l^* >\sqrt{-{\alpha^*}^2+2+2 \sqrt{-{\alpha^*}^2+1}}$, the central object is a traversable wormhole. Here we have defined the dimensionless quantities: $\l^*=\l / M, {\alpha^*}=\alpha / M$ for convenience.

 \section{$\text{Circular orbit on the equatorial plane in RSV spacetime}$}

In the spacetime described by equation (\ref{1}), the geodesic motion of particles near the central body is governed by $g_{\mu \nu} \dot{x}^\mu \dot{x}^\nu=\epsilon$. When $\epsilon=0$, it corresponds to a null geodesic, and when $\epsilon=-1$, a time-like geodesic is corresponding. For a massive particle moving along a time-like geodesic on the equatorial plane, its energy $E$, angular momentum $L$, and effective potential $V_{\text {eff }}$ are
\begin{equation}
\frac{d t}{d \tau}=\Sigma \frac{E A-2 L M \alpha x}{(2 M \alpha x)^2+(\Sigma-2 Mx) A}, \label{4}
\end{equation}
\begin{equation}
\frac{d \phi}{d \tau}=\Sigma \frac{2 E M \alpha x+L(\Sigma-2 Mx)}{(2 M \alpha x)^2+(\Sigma-2 Mx) A}, \label{5}
\end{equation}
\begin{equation}
V_{e f f}=-1+\Sigma \frac{E^2 A-2 E L 2 M \alpha x-L^2(\Sigma-2 Mx)}{(2 M \alpha x)^2+(\Sigma-2 Mx) A}, \label{6}
\end{equation}
with $x=\sqrt{r^2+l^2}$. One can get the energy and angular momentum of a particle orbiting a central body in a circular orbit through $V_{\text {eff }}(r)=0, d V_{\text {eff }}(r) / d r=0$
\begin{equation}
E=\frac{x+2 M\left(-1+\alpha \omega_0\right)}{x \sqrt{1-\left(x^2+\alpha^2\right) \omega_0^2-2 Mx^{-1}\left(-1+\alpha \omega_0\right)^2}}, \label{7}
\end{equation}
\begin{equation}
L=\frac{\left(x^2+\alpha^2\right) x \omega_0+2 M \alpha\left(-1+\alpha \omega_0\right)}{x \sqrt{1-\left(x^2+\alpha^2\right) \omega_0^2-2 Mx^{-1}\left(-1+\alpha \omega_0\right)^2}}. \label{8}
\end{equation}
Where $\omega_0$ is the angular velocity of the particle motion
\begin{equation}
\omega_0=\frac{d \phi}{dt}=\frac{Mr}{ \pm \sqrt{Mr^2 x^3}+Mr \alpha}. \label{9}
\end{equation}
Here $+/-$ represents the angular velocity of particles rotating in the positive or negative directions, respectively. To ensure that particles moving along timelike geodesics indeed have circular orbits, it is necessary to requre that equations (\ref{7}) and (\ref{8}) should have significance in physics. That is, $1-\left(x^2+\alpha^2\right) \omega_0^2-2 Mx^{-1}\left(-1+\alpha \omega_0\right)^2>0$ must be guaranteed. Substituting equation (\ref{9}) into this inequality, we obtain: for particles rotating in the positive $(+)$ or negative $(-)$ direction around a central body in RSV spacetime, if the circular orbits exist, the following expression needs to be satisfied
\begin{equation}
r x^2(-3 M+x) \pm 2 \sqrt{M r^2 x^3} \alpha>0. \label{10}
\end{equation}

Expression (\ref{10}) shows that the region of circular orbits is influenced by $M$, $\alpha$, $l$.
In Figure 1, we take particles rotating positively as an example and plot the existence domain of circular orbits as a function of the spin parameter $\alpha^*$ for particles moving near three different types of central bodies (e.g. Kerr black hole $l^*=0$, regular black hole $l^*=1$, and wormhole $l^*=3$ ) in the RSV spacetime. For convenience, we set $M=1$.

 \begin{figure}[ht]
\includegraphics[width=5.5cm]{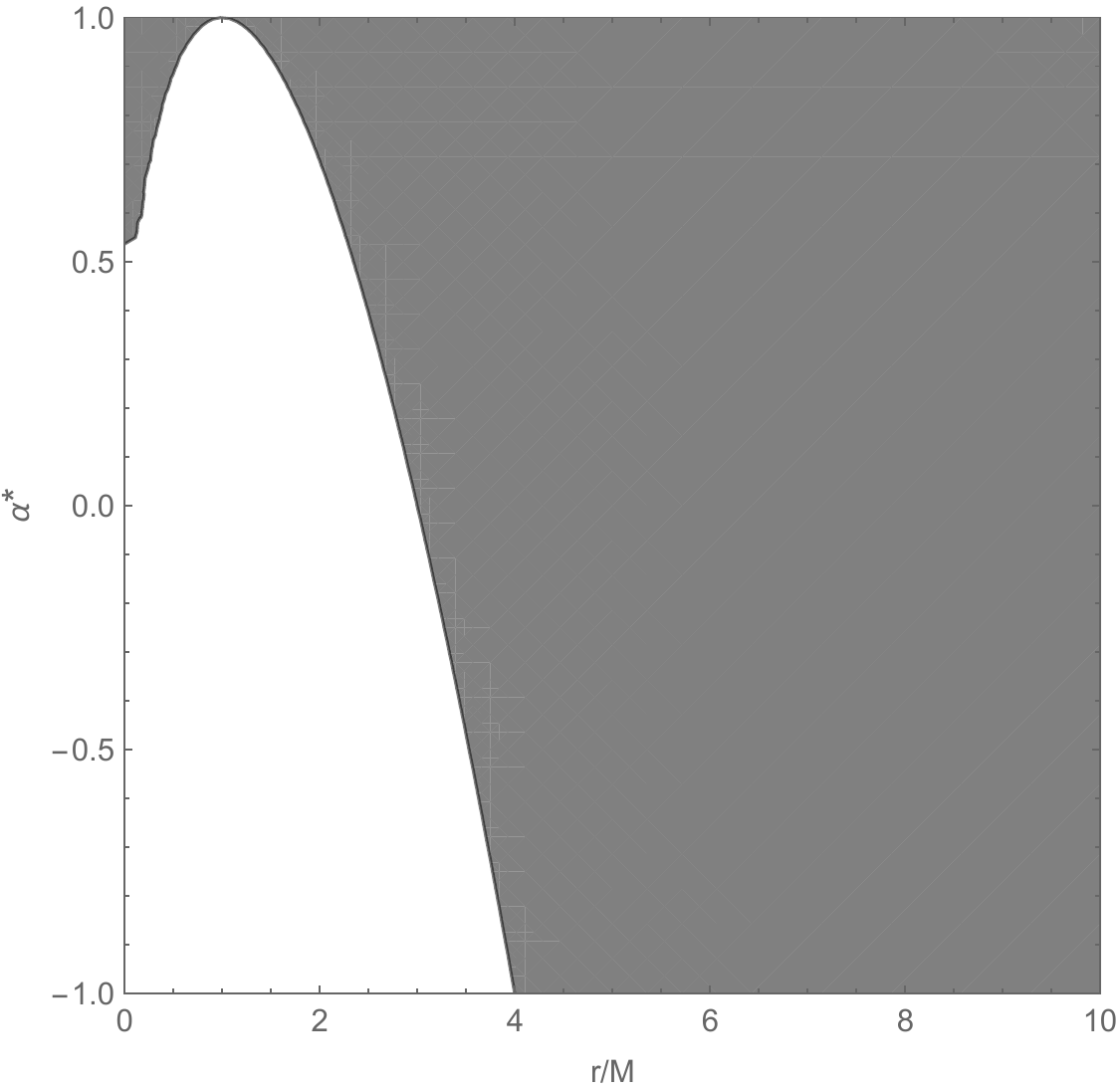}
\includegraphics[width=5.5cm]{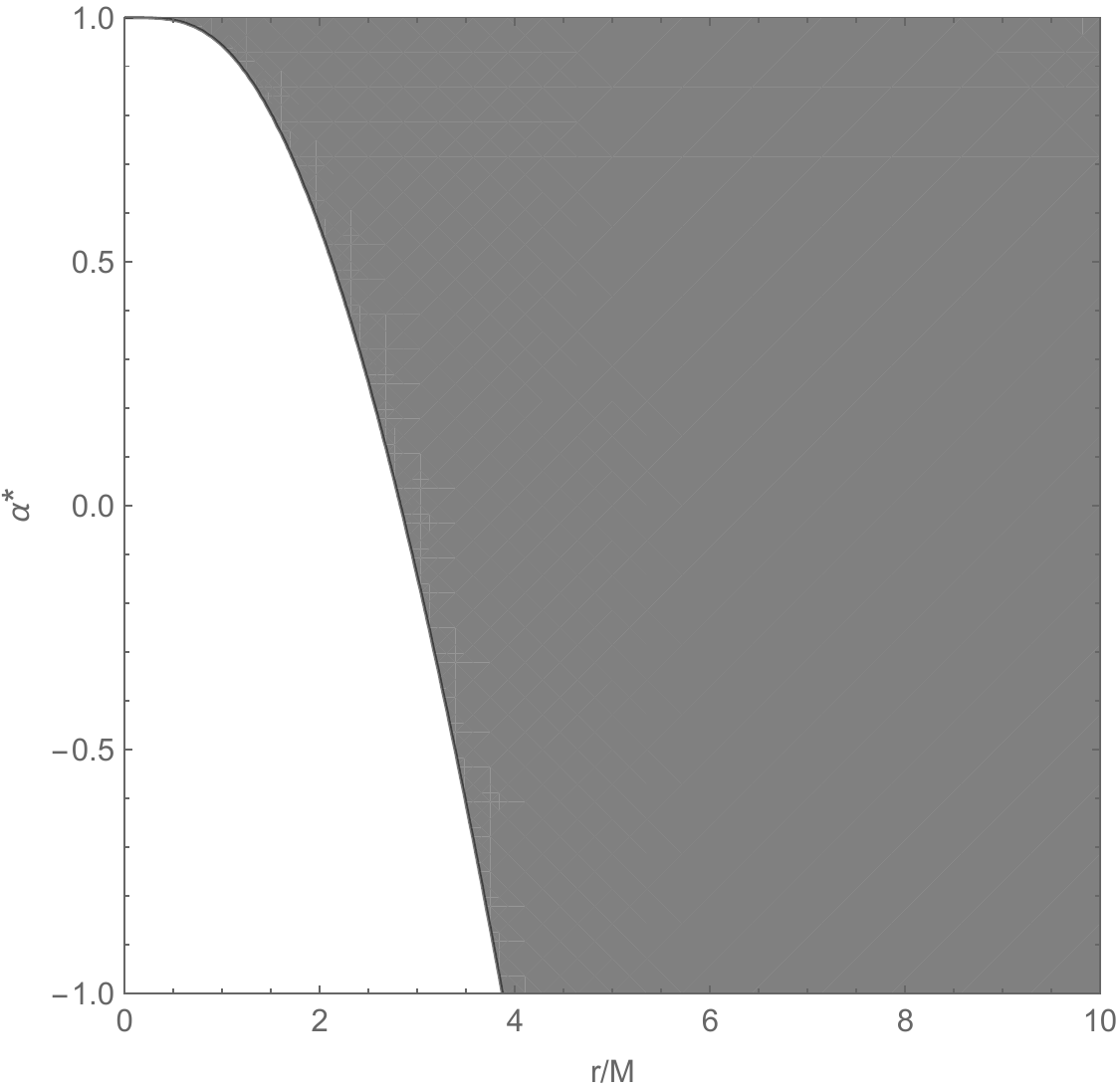}
\includegraphics[width=5.5cm]{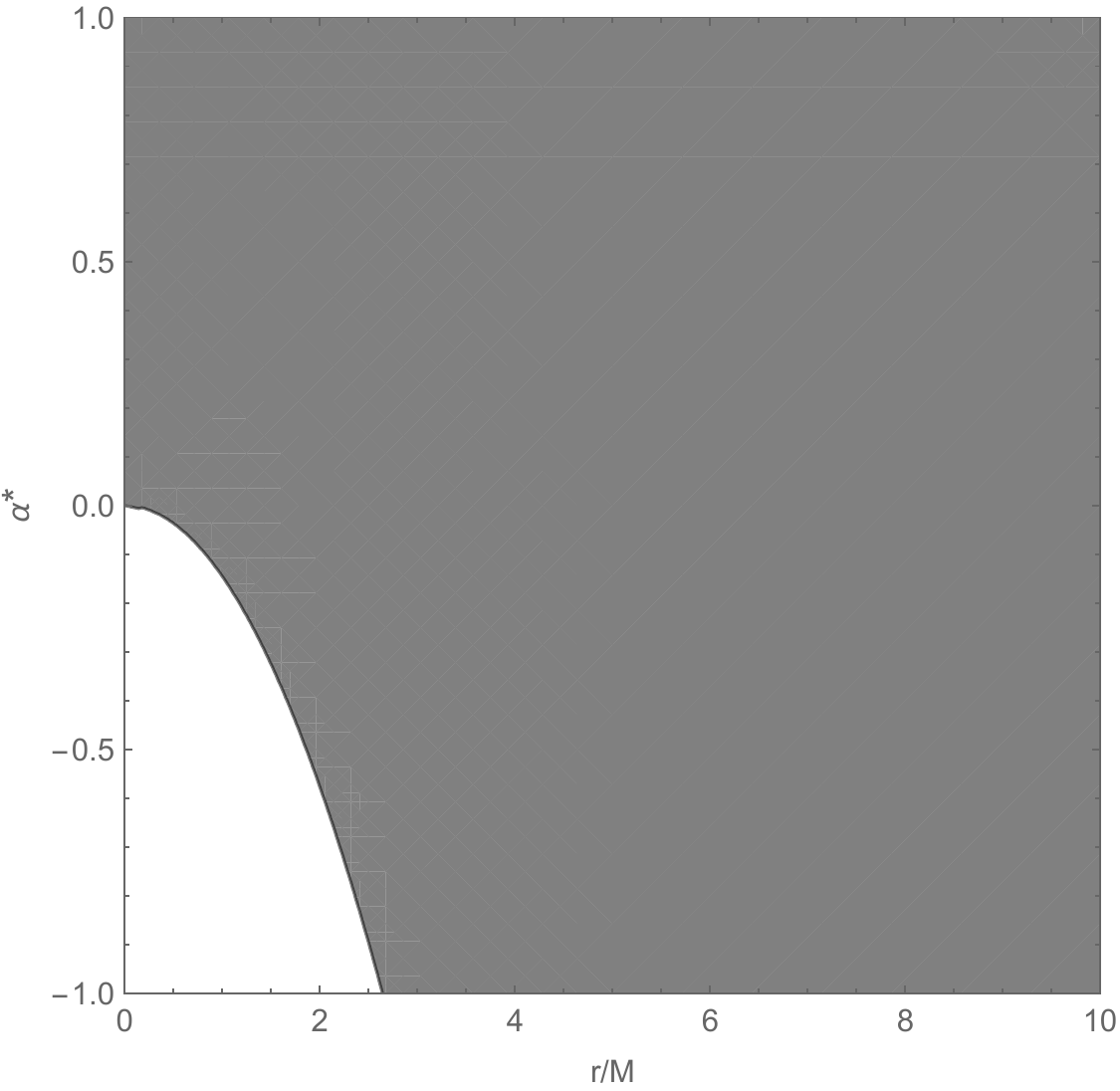}
  \caption{In RSV spacetime, the circular orbit of a particle exists in a region (shaded area), depending on the spin parameter $\alpha^*$. The values of parameter $l^*$ are taken to be 0, 1, and 3 from left to right, respectively.}\label{fig:1}
\end{figure}

From Figure \ref{fig:1}, we can see the possible circular orbit positions of particles orbiting axially symmetric celestial bodies in the RSV spacetime. Clearly, for the positively or negatively rotating Kerr black holes, the regular black holes, and the negatively rotating wormholes, the circular orbits need to be confined to specific locations (corresponding to the shaded regions depicted in the figure); for positively rotating wormholes, the circular orbit positions of particles can be located at anywhere in the region $r>0$.

\section{$\text{Stability of circular orbits of oscillatory particles in RSV spacetime}$}
To investigate the stability of circular orbits and obtain the angular frequency of the oscillatory motion of particles on stable circular orbits, we consider the geodesic equation \cite{59}
\begin{equation}
\frac{d^2 x^\mu}{d p^2}+\Gamma_{\alpha \beta}^\mu \frac{d x^\alpha}{d p} \frac{d x^\beta}{d p}=0. \label{11}
\end{equation}
$p$ is the affine parameter on the geodesic, and equation (\ref{11}) describes the motion of a particle orbiting a central body on the equatorial plane.
By using the successive approximation method, the motion of test particles can be analyzed through three fundamental frequencies \cite{epicyclic motion-0}. In this framework, the zeroth-order approximation describes the circular motion confined to the equatorial plane, characterized by the standard orbital frequency \cite{epicyclic motion-1}. Introducing small disturbances to circular orbits results in epicyclic motion, which, under the first-order approximation, manifests as two independent oscillations in the radial and vertical directions\cite{epicyclic motion-0,epicyclic motion-1,epicyclic motion-2, epicyclic motion-3, epicyclic motion-4, 92}.
Next, consider the epicyclic motion of a particle due to small perturbations at the circular orbit, and the deviation vector is introduced into the geodesic equation, where
$Z^\mu(p)$ describes the circular unperturbed motion. Then for a circular orbit in the equatorial plane, the deviation vector is written as
\begin{equation}
\xi^\mu(p)=x^\mu(p)-Z^\mu(p), \label{12}
\end{equation}
with the coordinates of the particle
\begin{equation}
Z^\mu(p)=\left\{t(p), r_0, \pi / 2, \omega_0 t(p)\right\}. \label{13}
\end{equation}
Here $t(p)$ represents the time coordinate as a function of the affine parameter
$p$; $r_0$ represents the radial coordinate position of the particle when it orbits around the central body in a circular orbit; The particle remains in the equatorial plane, indicated by $\pi/2$. Additionally, the azimuthal angle $\omega_0 t(p)$ evolves linearly with time, representing uniform angular velocity motion. Substituting equation (\ref{12}), (\ref{13}) into (\ref{11}), expanding $\xi^\mu(p)$ in its power series form, and taking the first-order linear approximation, we obtain a deviation equation for $\xi^\mu(p)$. The equation describes the motion of a particle on a circular orbit with the small perturbation \cite{92}
\begin{equation}
\begin{gathered}
\frac{d^2 \xi^\mu}{d t^2}+2 \gamma_\sigma^\mu \frac{d \xi^\sigma}{d t}+\xi^a \partial_a U^\mu=0, \quad a \equiv r, \theta, \\
\gamma_\sigma^\mu=\Gamma_{\sigma \beta}^\mu u^\beta\left(u^0\right)^{-1}, U^\mu=\gamma_\sigma^\mu u^\sigma\left(u^0\right)^{-1}.
\end{gathered} \label{14}
\end{equation}
The quantities $\gamma_\sigma^\mu$ and $\partial_a U^\mu$ are set on a circular orbit $r=r_0, \theta=\pi / 2, u^\mu=\dot{Z^\mu}=u^{0}(1, 0, 0, \omega_{0})$. When $\mu=\mathrm{r}, \theta$, two decoupling oscillations in the radial and vertical directions can be obtained
\begin{equation}
\frac{d^2 \xi^r}{d t^2}+\omega_r^2 \xi^r=0, \frac{d^2 \xi^\theta}{d t^2}+\omega_\theta^2 \xi^\theta=0. \label{15}
\end{equation}
Through equation (\ref{14}), (\ref{15}), the following is obtained
\begin{equation}
\omega_r^2=\partial_r U^r-4 \gamma_A^r \gamma_r^A, \omega_\theta^2=\partial_\theta U^\theta. \label{16}
\end{equation}
Substitute equations (\ref{13}) and (\ref{14}) into equation (\ref{16}) and expand it, the general expressions for $\omega_r^2$ and $\omega_\theta^2$ can be obtained as
\begin{equation}
\begin{aligned}
\omega_r^2=\partial_r \Gamma_{t t}^r+2 \omega_0 \partial_r \Gamma_{t \phi}^r & +\omega_0^2 \Gamma_{\phi \phi}^r-4\left(\Gamma_{t t}^r \Gamma_{r t}^t+\Gamma_{t t}^r \Gamma_{r \phi}^t \omega_0+\Gamma_{t \phi}^r \Gamma_{r t}^t \omega_0+\Gamma_{t \phi}^r \Gamma_{r \phi}^t \omega_0^2+\Gamma_{\phi t}^r \Gamma_{r t}^\phi\right. \\
& \left.+\Gamma_{\phi t}^r \Gamma_{r \phi}^\phi \omega_0+\Gamma_{\phi \phi}^r \Gamma_{r t}^\phi \omega_0+\Gamma_{\phi \phi}^r \Gamma_{r \phi}^\phi \omega_0^2\right),
\end{aligned} \label{16.1}
\end{equation}
\begin{equation}
\omega_\theta^2=\partial_\theta \Gamma_{t t}^\theta+2 \omega_0 \partial_\theta \Gamma_{t \phi}^\theta+\omega_0^2 \Gamma_{\phi \phi}^\theta. \label{16.2}
\end{equation}

When $\omega_r^2 \geq0$ and $\omega_\theta^2 \geq0$, the two perturbation quantities $\xi_r$ and $\xi_\theta$ in equation (\ref{15}) exhibit oscillatory behavior with respect to time $t$. The particle undergoes simple harmonic oscillations in the radial and vertical directions along the circular orbit with frequencies $\omega_r$ and $\omega_\theta$, respectively, which are referred to as epicyclic frequencies. Under the linear approximation, the circular motion is stable.
When $\omega_r^2 < 0$ or $\omega_\theta^2 < 0$, the solution $\xi_r$ or $\xi_\theta$ of equation (\ref{15}) varies exponentially with respect to time $t$, indicating that small perturbations in the corresponding direction will deviate exponentially from the circular orbit, leading to instability in the motion \cite{59,epicyclic motion-0}. In the RSV spacetime, using expression (\ref{16.1}), (\ref{16.2}), we obtain expressions for $\omega_r^2$ and $\omega_\theta^2$ as follows

\begin{equation}
\omega_\theta^2=\frac{1}{x^5}\left[2 M \alpha^2-4 M \omega_0 \alpha\left(x^2+\alpha^2\right)+\omega_0^2\left(x^5+x^2 \alpha^2(4 M+x)+2 M \alpha^4\right)\right], \label{17}
\end{equation}
\begin{equation}
\begin{aligned}
\omega_r^2=\frac{1}{x^7}\left\{\omega_0^2[\right. & x^4\left(l^2(2 M-x)+r^2(3 x-8 M)\right)+x \alpha^2\left(-l^4+r^4+2 M r^2(M-5 x)\right. \\
& \left.\left.+l^2 M(x-2 M)\right)+M \alpha^4\left(l^2-4 r^2\right)\right]+2 M \alpha \omega_0\left[-l^4+l^2\left(5 r^2+2 M x\right.\right. \\
& \left.\left.-\alpha^2\right)+2 r\left(3 r^2-M x+2 \alpha^2\right)\right]+M\left[l^4+l^2\left(-r^2-2 M x+\alpha^2\right)\right. \\
& \left.-2 r^2\left(r^2-M x+2 \alpha^2\right)\right].
\end{aligned} \label{18}
\end{equation}

\begin{figure}[ht]
\includegraphics[width=8.5cm]{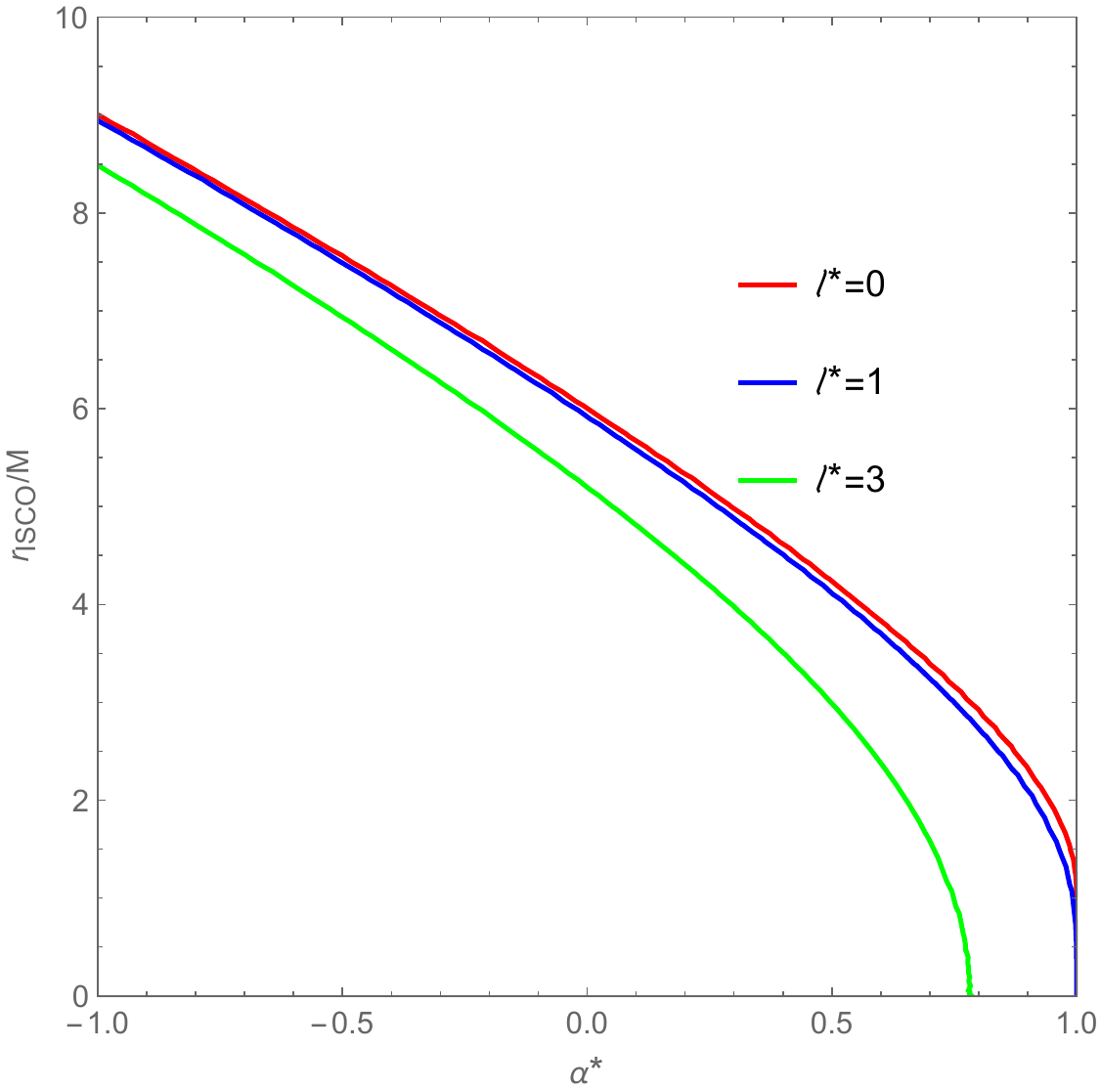}
  \caption{ISCO positions of different types of celestial bodies as a function of spin parameter $\alpha^*$.}\label{fig:2}
\end{figure}

Equations (\ref{17}) and (\ref{18}), show that the vertical epicyclic motion of particle $\omega_\theta^2>$ 0 is always stable. Therefore, the stability of particle's motion around the central celestial body is only controlled by $\omega_r^2(r)$. When $\omega_r^2(r)=0$, it corresponds to the position of the ISCO. In Figure \ref{fig:2} we show the curves of the ISCO position for three different types of celestial bodies in RSV spacetime as a function of the spin parameter $\alpha^*$, where the red curve corresponds to the Kerr black hole $(l^*=0)$, the blue curve to the regular black hole (e.g. $l^*=1$), and the green curve to the traversable wormhole (e.g. $l^*=3$).

 We can see from Figure \ref{fig:2}, for axisymmetric SV objects that rotate in the positive direction and for static SV objects: $\alpha^* \in[0,1]$, the location of the ISCO will gradually approach the center of the radial coordinate as $\alpha^{*}$ increases. In contrast, for the RSV objects with rotation in the opposite direction, $\alpha^* \epsilon[-1,0)$, the position of ISCO moves further away from the radial coordinate center as $|\alpha^*|$ value increases. For both positive and negative rotation of the RSV objects, the radius of the ISCO decreases as the parameter $l^*$ increases. When the spin parameter value is fixed, the location of ISCO is close to the radial coordinate center with the increase of $l^*$, which means that the traversable wormhole spacetime described by equation (\ref{1}) has a stable particle trajectory closer to the central object than the Kerr black hole and the rotating regular black hole.

\begin{figure}[ht]
\includegraphics[width=5.5cm]{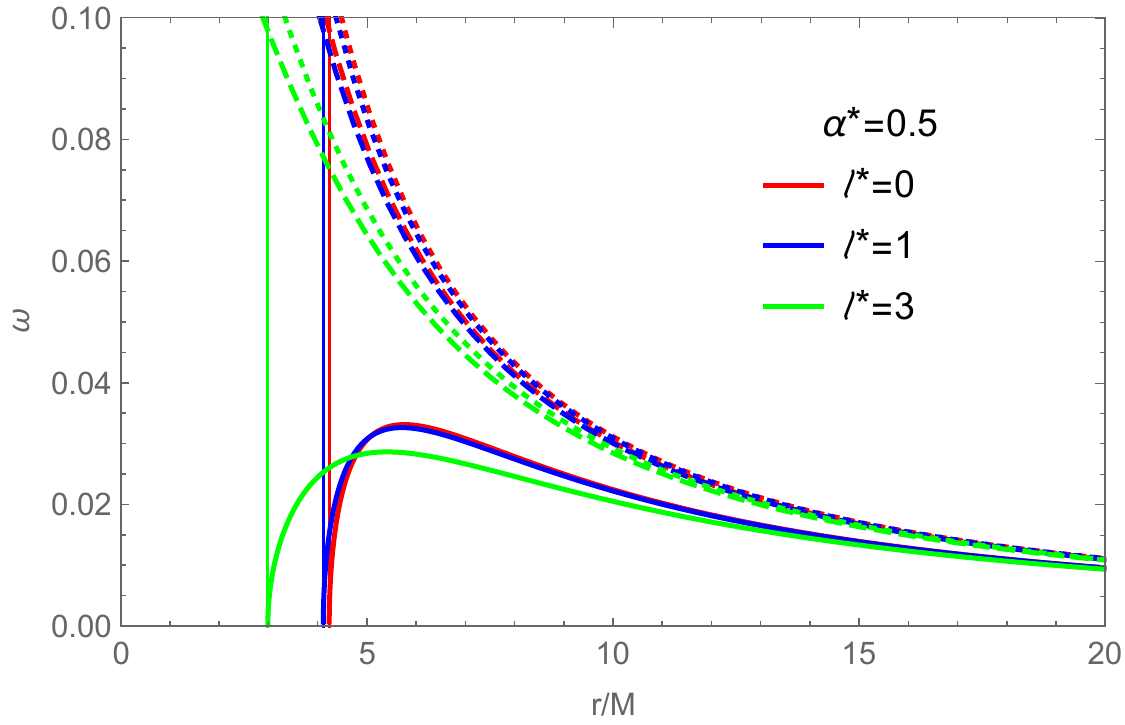}
\includegraphics[width=5.5cm]{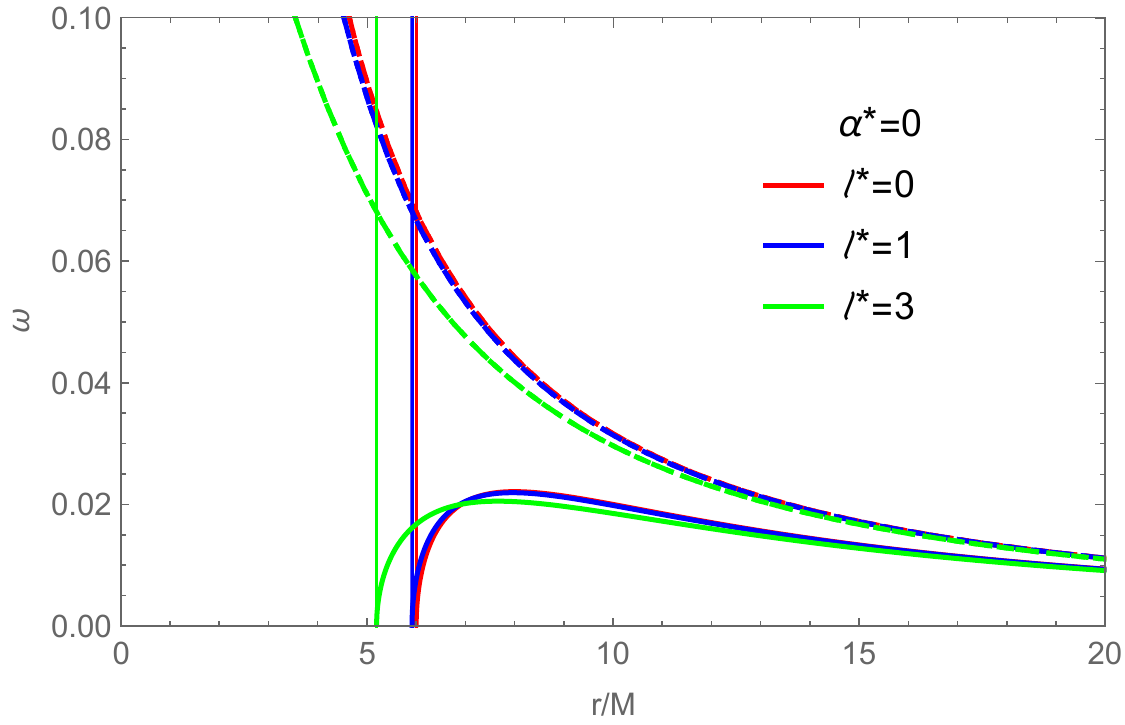}
\includegraphics[width=5.5cm]{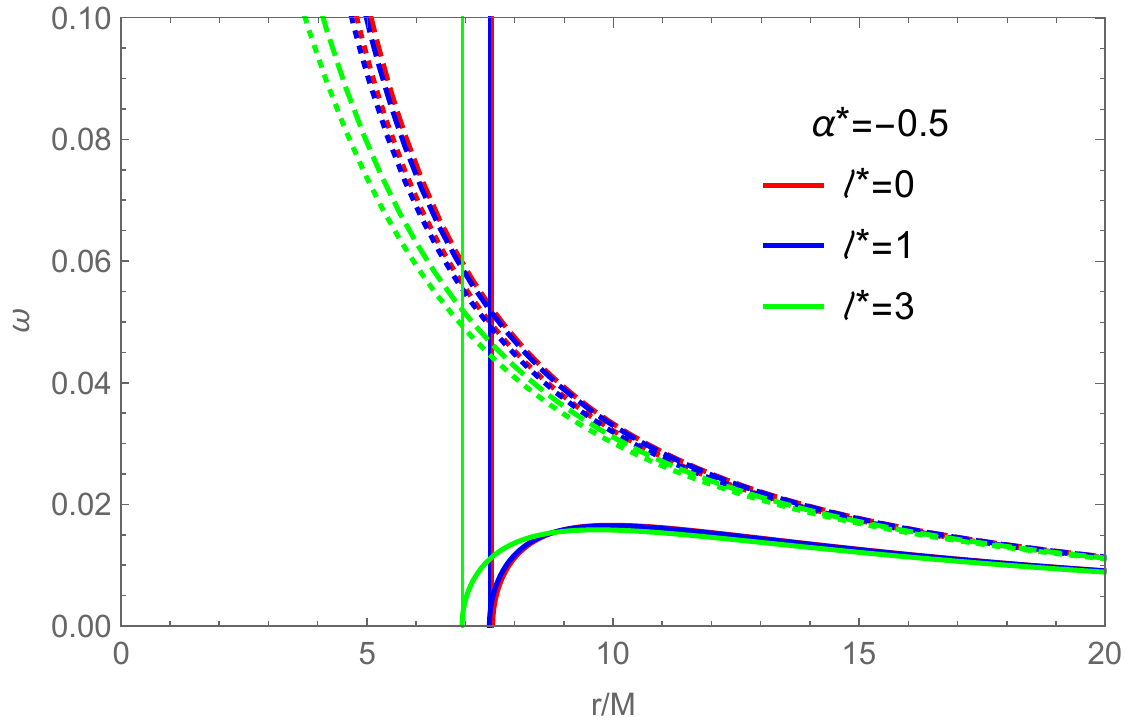}
  \caption{Angular frequencies $\omega_\theta$ (dashed line), $\omega_r$ (solid line), and $\omega_0$ (dotted line) of the oscillating motion of particles around three types of the RSV celestial bodies that are in forward rotation, at rest, and in reverse rotation, as a function of radial coordinate $r/M$. The values of $\alpha^*$ from left to right are 0.5, 0, and -0.5 . The vertical line indicates the location of the $r_{\text {ISCO }}/M$.}\label{fig:3}
\end{figure}

ISCO plays an important role in astrophysics, as it represents the inner boundary of the Keplerian accretion disk around celestial bodies and determines the starting position of the accretion disk \cite{94,95,96,97}. It is generally assumed that the epicyclic motion of particles around the central celestial body is probably generated by the motion of the accretion flow within the accretion disk around the celestial body \cite{62,78}. To further analyze the nature of the oscillatory motion of particles around the celestial bodies, the angular frequencies of the three types of the RSV celestial bodies, for cases of positive rotation $(\alpha^*=0.5)$, stationary $(\alpha^*=0)$, and negative rotation $(\alpha^*=-$ 0.5 ), are plotted in Figure \ref{fig:3}. The values of $\omega_\theta$ (dashed line), $\omega_r$ (solid line), and $\omega_0$ (dotted line) are plotted as a function of radial coordinate $r/M$.

When $r<r_{\text {ISCO}}$, the oscillating motion of the particle is unstable to radial frequency perturbations. Therefore, in this article we only study the case of $r>r_{\text {ISCO}}$. Figure \ref{fig:3} shows that Kerr black holes, regular black holes, and traversable wormholes always have $\omega_\theta\left(\omega_0\right)>\omega_r$, when spin parameter $\alpha^*=0 \pm 0.5$. Specifically, for the RSV objects $(\alpha^*=0.5)$ that rotate in the positive direction, $\omega_0>\omega_\theta>$ $\boldsymbol{\omega}_r$. For the RSV objects $(\alpha^*=-0.5)$ in the negative direction, $\omega_\theta>$ $\omega_0>\omega_r$. When $\alpha^*=0$, it is reduced to the case of spherically symmetric SV spacetime, where $\omega_0=\omega_\theta>\omega_r$. At the same time, Figure \ref{fig:3} shows that if the central object is a traversable wormhole, the frequency value of the particles orbiting it is lower than that of regular black holes and Kerr black holes. The primary source of the QPOs phenomenon is considered to be the particle's orbital motion around the central object, and models can be constructed to study the behavior of QPOs in celestial objects \cite{98}.

\section{$\text{Research on quasi periodic oscillation in RSV spacetime}$}
A high probability that the ratio of high and low frequency $\left(v_u: v_l\right)$ of HFQPOs twin peaks occurs in a fixed ratio of 3:2 is observed in low-mass X-ray binaries containing black holes, as indicated by many observations \cite{99}. This suggests that there might be a resonance within the accretion disk surrounding the central object. The frequency of the observed HFQPO features was found to be very close to the frequency of test particles that oscillate near the star's ISCO, thus the HFQPO phenomenon is considered to be a manifestation of strong gravitational effects. Due to the high precision of HFQPOs, it is now generally accepted that they are an effective means of testing gravitational theories. To date, the study of the HFQPO phenomenon has mostly involved the construction of a linear combination between the radial, latitudinal, and orbital frequencies of particles performing epicyclic motion around the central celestial body \cite{100,101}.

The twin peaks HFQPO models can be roughly divided into four categories \cite{101}: tidal precession model \cite{102}, disko-seismic model \cite{103,104}, relativistic precession model \cite{105,106}, and resonance model \cite{80}. However, so far, no model has been able to simultaneously account for the observations from different sources in the black hole system within the GR framework, making it impossible to determine the exact physical mechanism responsible for the HFQPO phenomenon \cite{76,77,78,79}. This may suggest the need to incorporate more complex QPO models, such as those accounting for the influence of magnetic fields \cite{cichang-1,cichang-2}, or indicate that the observed frequencies might not correct HFQPOs \cite{78}.
Although we can still use various models to analyze the QPOs around different central bodies and obtain information about the central object and the accretion disk surrounding it, such as the possible resonant modes, the location of resonance occurrence, the peak frequency \cite{59}, and the range of model parameters in gravitational theory \cite{42}. However, obtaining such information strongly depends on the selection of the HFQPO models. Therefore, in this section, we consider several popular HFQPO models, including ER $
\left(\mathrm{ER}_0, \mathrm{ER}_1, \mathrm{ER}_2, \mathrm{ER}_3, \mathrm{ER}_4, \mathrm{ER}_5\right)
$ \cite{80,81}, RP $\left(\mathrm{RP}_0, \mathrm{RP}_1, \mathrm{RP}_2\right)$ \cite{82,83}, TD \cite{84,85,86}, and WD \cite{87}. We investigate the possible values of the twin peak frequency of oscillating particles around the central object in different models in the RSV spacetime as well as the location of the 3:2 resonance. To make the physical quantities in the theoretical model consistent with the dimensions of corresponding observables, we define
\begin{equation}
v=\frac{1}{2 \pi} \frac{c^3}{G M} \omega_x, (x=r, \theta, 0). \label{19}
\end{equation}
Where $c$ is the speed of light, $G$ is the Newtonian gravitational constant, and $M$ is the mass of the celestial body.

Since the 3:2 resonance phenomenon may originate from the resonance between different oscillation modes within the accretion disk \cite{81,107}, ER model has been proposed to study the resonance of axisymmetric oscillation modes in the accretion disk \cite{accretion disk-1,accretion disk-2}. From a geometrical perspective, accretion disks can be classified into thin disks, thick disks, and toroidal accretion disks. For geometrically thin disks and elongated annular disks, the oscillation frequencies of the disk are associated with the orbital and epicyclic frequencies of circular geodesic motion \cite{accretion disk-3}. The ER model typically includes the parametric resonance model (PRM) and the forced resonance model (FRM), among others. Studies indicate that systems with thin disks or those approximating Keplerian disks are more likely to exhibit PRM characteristics \cite{81,107}, with PRM dynamics governed by the Mathieu equation. Additionally, considering the potential dissipation, pressure effects, and other physical influences within the accretion disk \cite{accretion disk-4,accretion disk-5, accretion disk-6}, it becomes necessary to introduce forcing terms into the perturbation equation (\ref{15}) when analyzing the motion of test particles in the equatorial plane
\begin{equation}
\delta \ddot{\mathrm{r}}+v_{\mathrm{r}}^2 \delta \mathrm{r}=v_{\mathrm{r}}^2\mathrm{~F}_{\mathrm{r}}(\delta \mathrm{r}, \delta \theta, \delta \dot{r}, \delta \dot{\theta}), \delta \ddot{\theta}+v_\theta^2 \delta \theta=v_\theta^2\mathrm{~F}_\theta(\delta \mathrm{r}, \delta \theta, \delta \dot{\mathrm{r}}, \delta \dot{\theta}). \label{19.1}
\end{equation}
In this context, $F_r$ and $F_\theta$ denote two unspecified functions that account for the coupling effects arising from the perturbation terms. Within the framework of the PRM, it is assumed that $F_r=0$ and $F_\theta=h \delta \theta \delta r$, where $h$ is a constant \cite{accretion disk-7}. Under these assumptions, equation (\ref{19.1}) reduces to
\begin{equation}
\delta \ddot{\mathrm{r}}+v_{\mathrm{r}}^2 \delta \mathrm{r}=0, \delta \ddot{\theta}+v_\theta^2\left[1+\mathrm{h} \cos \left(v_{\mathrm{r}} \mathrm{t}\right)\right] \delta \theta=0, \label{19.2}
\end{equation}
and is excited when \cite{107,accretion disk-5,accretion disk-7,accretion disk-8}
\begin{equation}
\frac{v_r}{v_\theta}=\frac{2}{n},(n=3,4,5\ldots) . \label{19.3}
\end{equation}
It is generally that frequencies corresponding to lower-order resonances are preferred, as they result in a larger amplitudes of the observed signal \cite{59}.

In a more practical scenario, the presence of pressure, viscosity, or magnetic stresses within the accretion flow can give rise to non-zero forcing terms \cite{accretion disk-4,accretion disk-9}. Building upon this idea, the FRM was introduced. Within this framework, the perturbation equation incorporating non-zero forcing terms takes the following form
\begin{equation}
\delta \ddot{\theta}+v_\theta^2 \delta \theta=-v_\theta^2 \delta r \delta \theta+\mathrm{F}_\theta(\delta \theta). \label{19.4}
\end{equation}
Where $\delta r=B \cos \left(v_r t\right), F_\theta$ denotes the nonlinear term related to $\delta \theta$. The forced resonance is activated when the epicyclic frequency ratio satisfies the following relationship
\begin{equation}
\frac{v_\theta}{v_r}=\frac{p}{q}. \label{19.5}
\end{equation}
In the equation mentioned above, $p$ and $q$ are integers with relatively small values. Within the framework of resonant solutions, forced nonlinear resonance allows for the presence of combinational or beat frequencies \cite{accretion disk-10}. Moving forward, we consider that the observed HFQPO phenomena in these compact objects can be characterized within the RSV spacetime. We then explore how the previously discussed ER models account for the 3:2 twin peak frequency ratio observed in the X-ray emissions of such objects. In the PRM, the minimum value of the resonance parameter is $n=3$ $\left(v_\theta: v_r=3: 2 \;\left(E R_0\right)\right)$; In FRM, it is required that $p / q$ is always greater than 1 (e.g. the lower order resonance $p: q= (3: 1\;(E R_1)$, and $2: 1\;(E R_3))$. In addition, considering the existence of other combined frequencies, there are $p: q= 5: 2\; \left(E R_2\right), \;5: 1 \;\left(E R_4\right), \;5: 3\; \left(E R_5\right)$, and the amplitude of oscillations is decreasing with increasing $p, q$. Below, we summarize these common ER models in Table \ref{table1}.

\begin{table}[ht]
\begin{center}
\begin{tabular}{|c|c|c|c|c|c|c|}
\hline Model & $\mathrm{ER}_0$ & $\mathrm{ER}_1$ & $\mathrm{ER}_2$ & $\mathrm{ER}_3$ & $\mathrm{ER}_4$ & $\mathrm{ER}_5$ \\
\hline$v_u$ & $v_\theta$ & $v_\theta$ & $v_\theta-v_r$ & $v_\theta+v_r$ & $v_\theta+v_r$ & $v_r$ \\
\hline$v_l$ & $v_r$ & $v_\theta-v_r$ & $v_r$ & $v_\theta$ & $v_\theta-v_r$ & $v_\theta-v_r$ \\
\hline
\end{tabular}
\end{center}
\caption{\label{table1}
 ER model and its variants.}
\end{table}

\begin{figure}[ht]
\includegraphics[width=5.5cm]{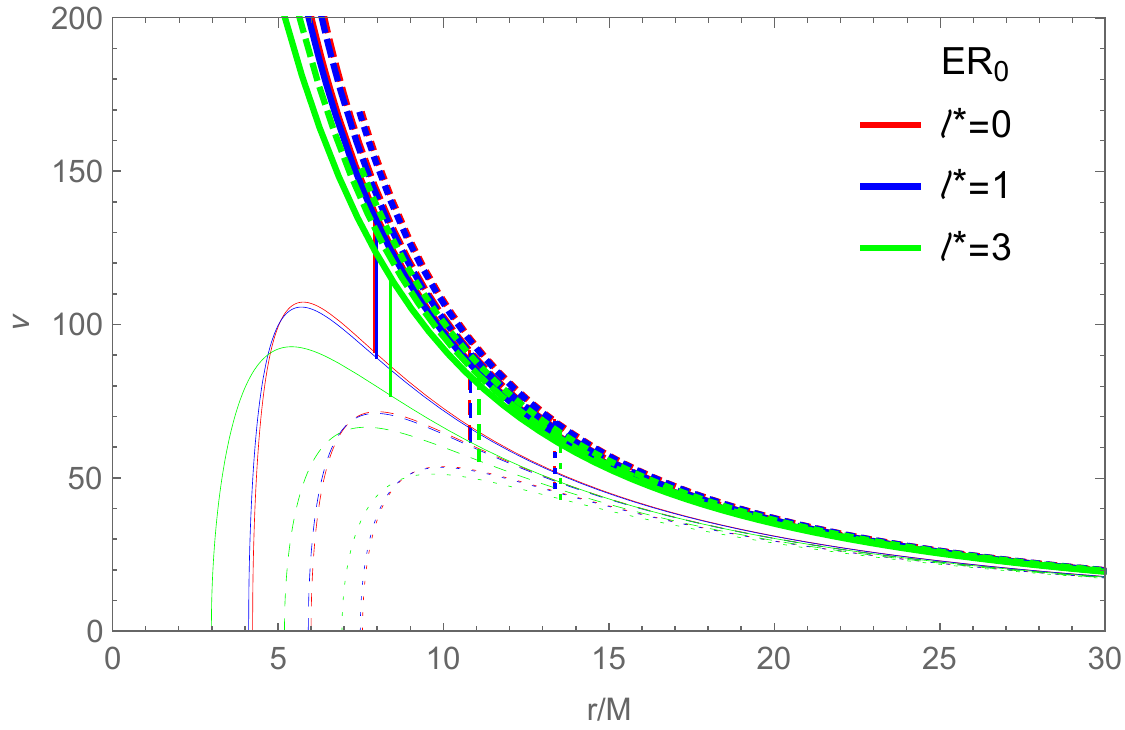}
\includegraphics[width=5.5cm]{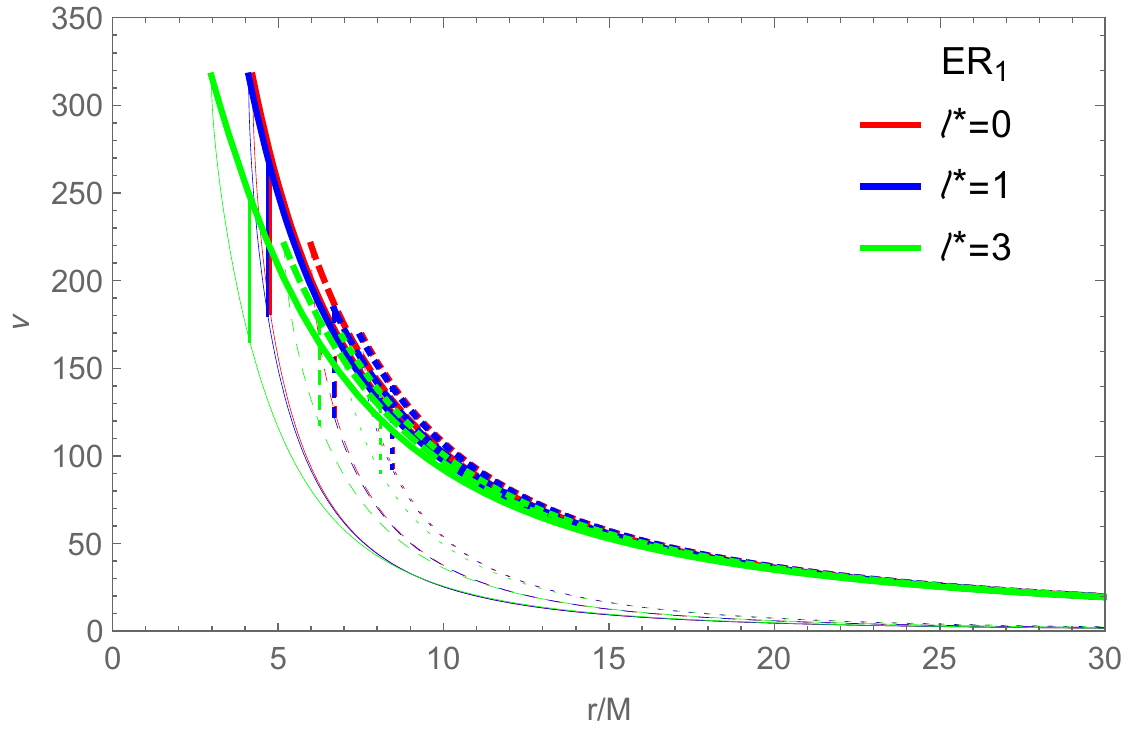}
\includegraphics[width=5.5cm]{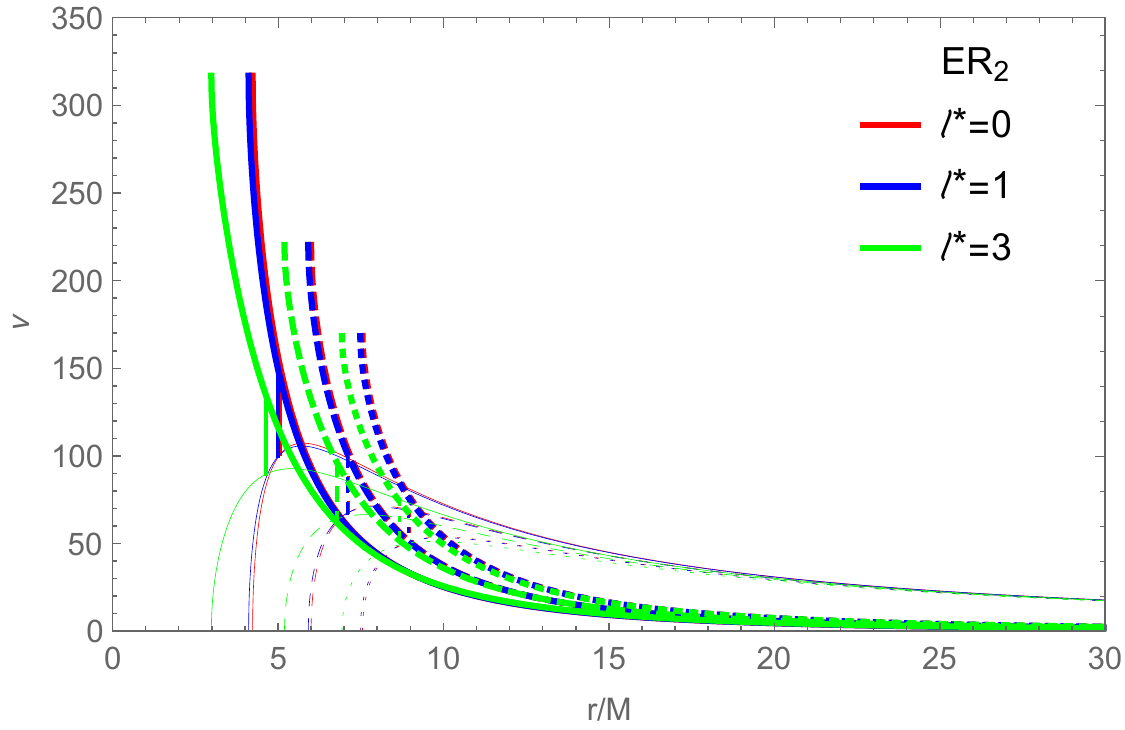}
\includegraphics[width=5.5cm]{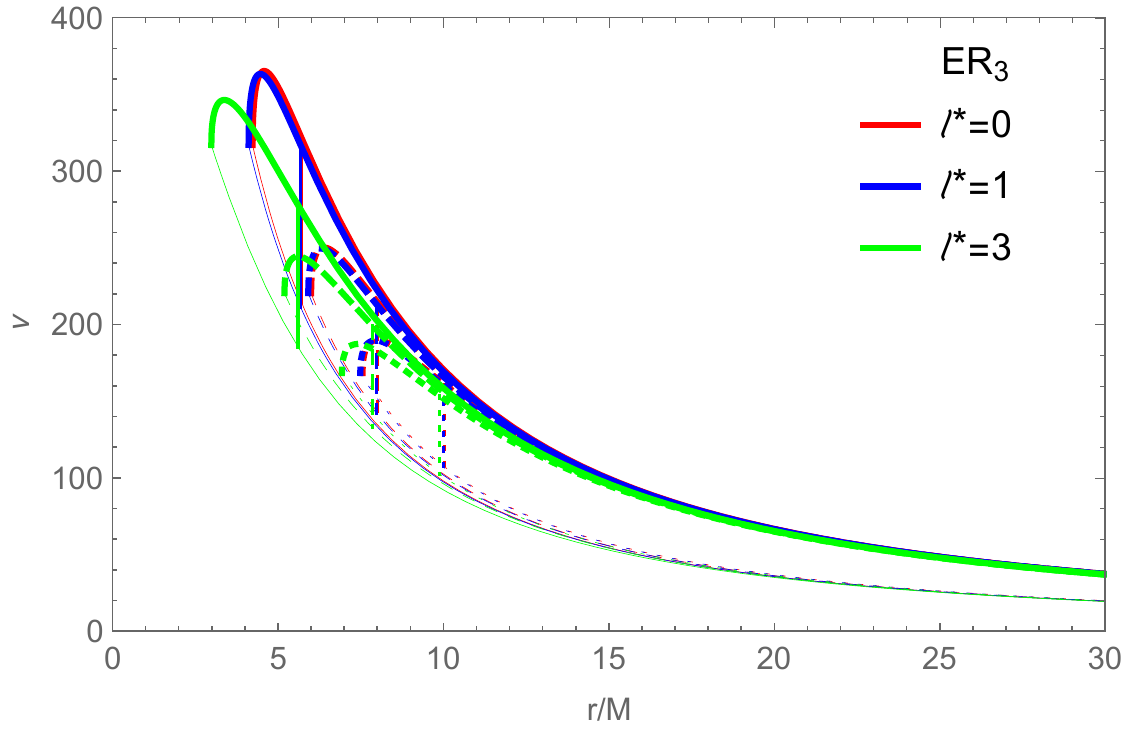}
\includegraphics[width=5.5cm]{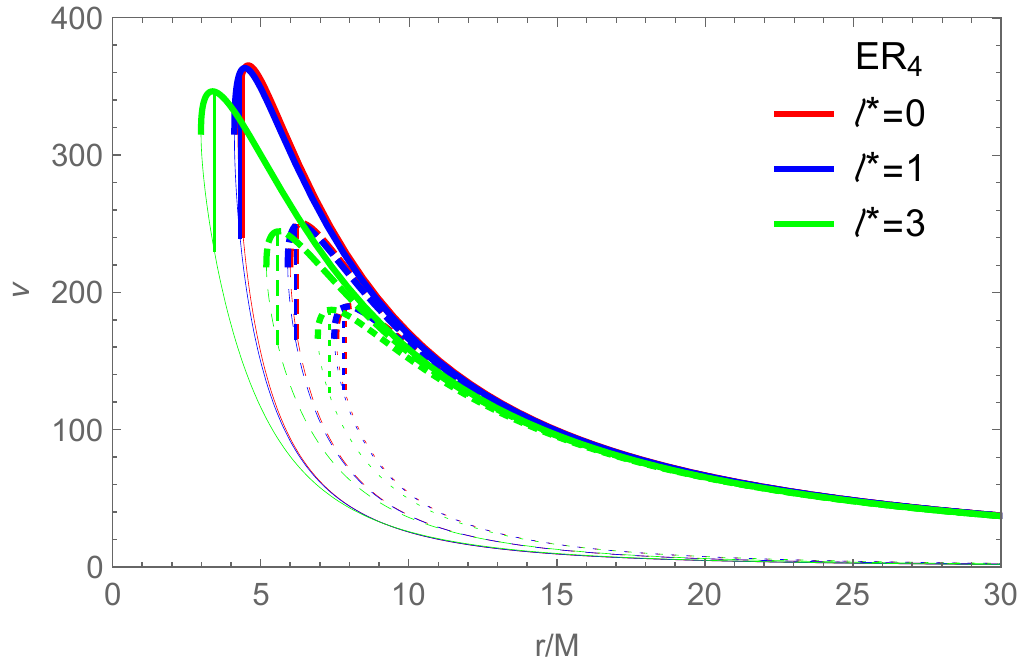}
\includegraphics[width=5.5cm]{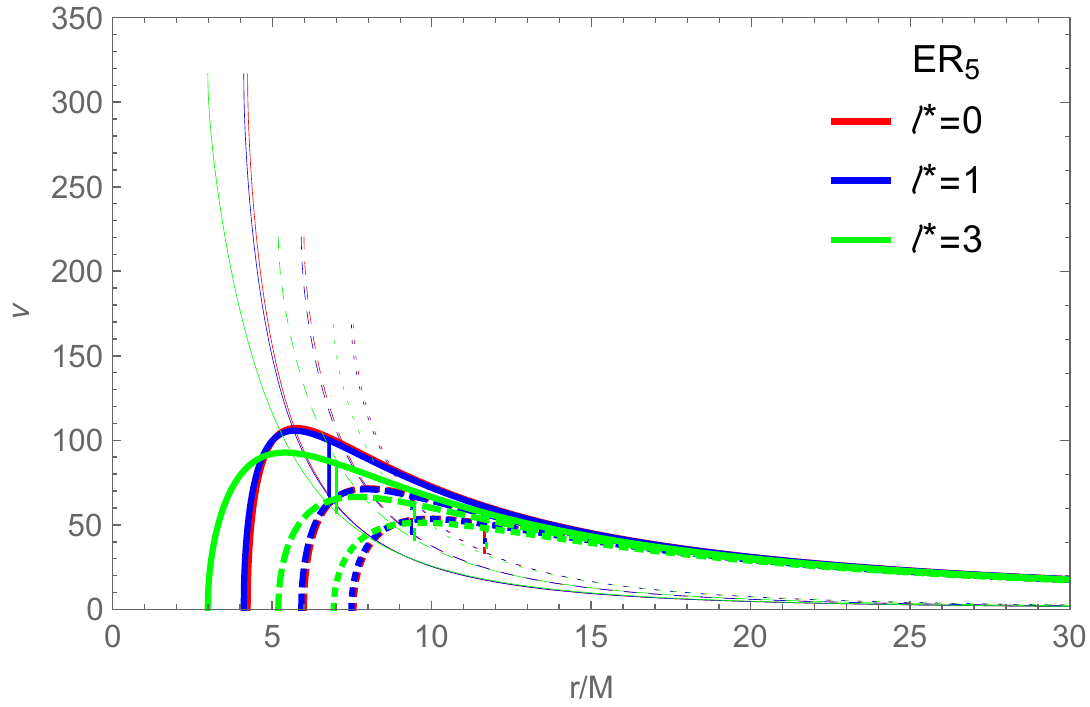}
  \caption{In the standard ER model and its variants, the particle undergoes epicyclic motion around three types of RSV celestial bodies with spin parameters $\alpha^*=0.5$ (solid line), $\alpha^*=0$ (dashed line), and $\alpha^*=-0.5$ (dotted line). The curves show the variation of the twin peaks frequency $v_u$ (thick line) and $v_l$ (thin line) with the radial coordinate $r/M$. The vertical lines represent the positions of the frequency ratio $v_u: v_l=3: 2$.}\label{fig:4}
\end{figure}

\begin{figure}[ht]
\includegraphics[width=5.5cm]{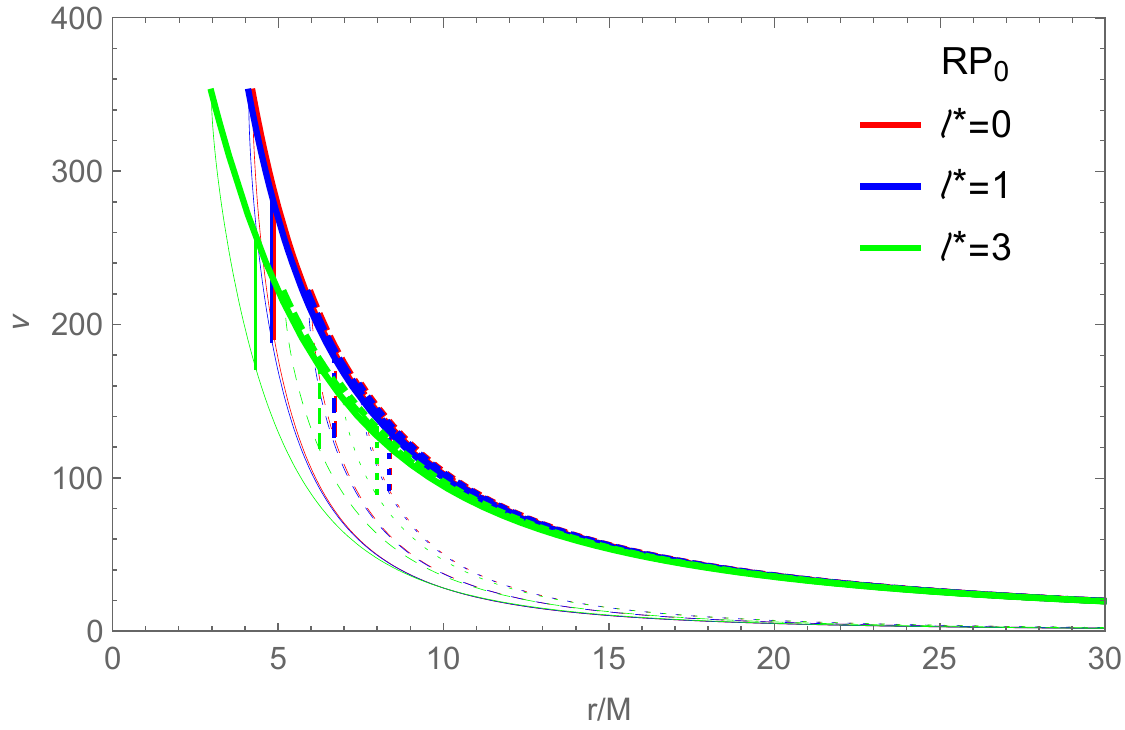}
\includegraphics[width=5.5cm]{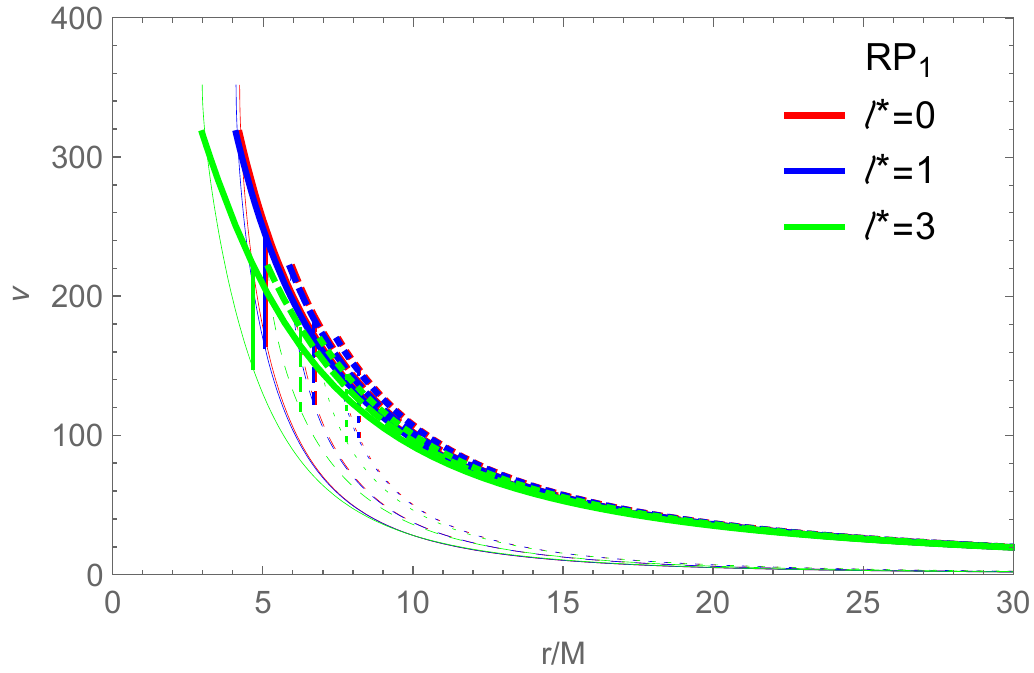}
\includegraphics[width=5.5cm]{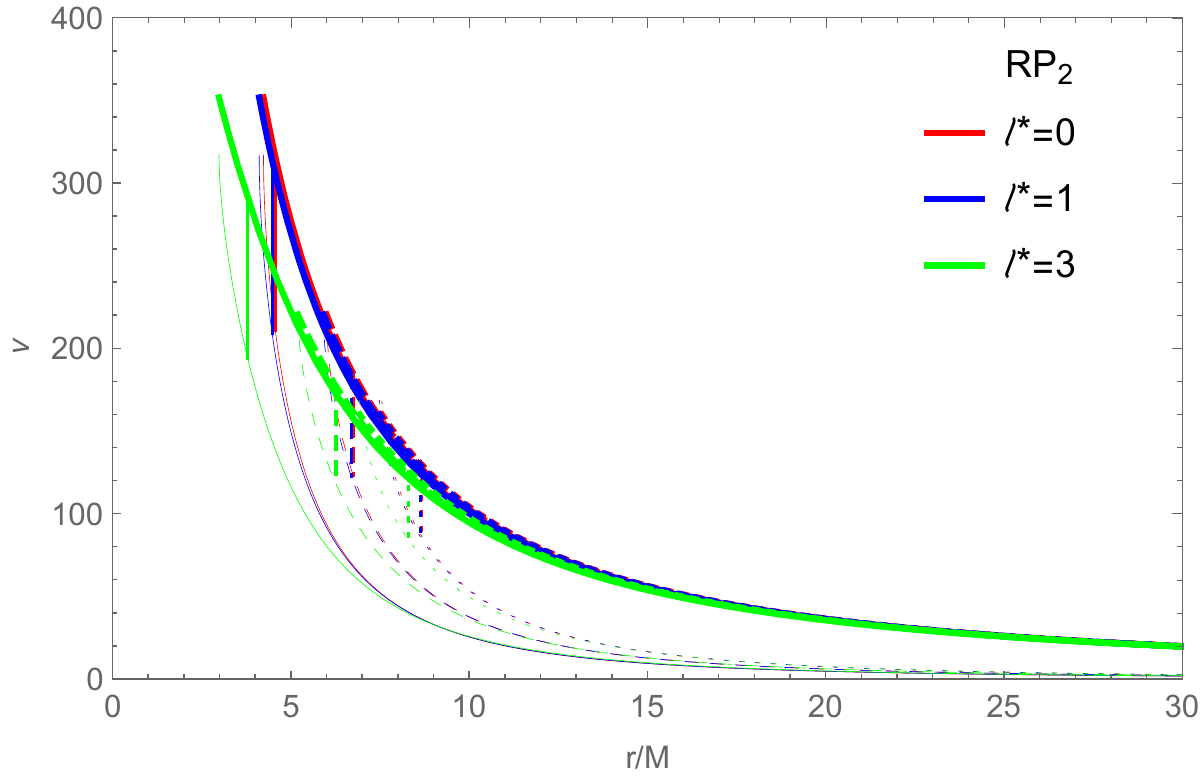}
\includegraphics[width=5.5cm]{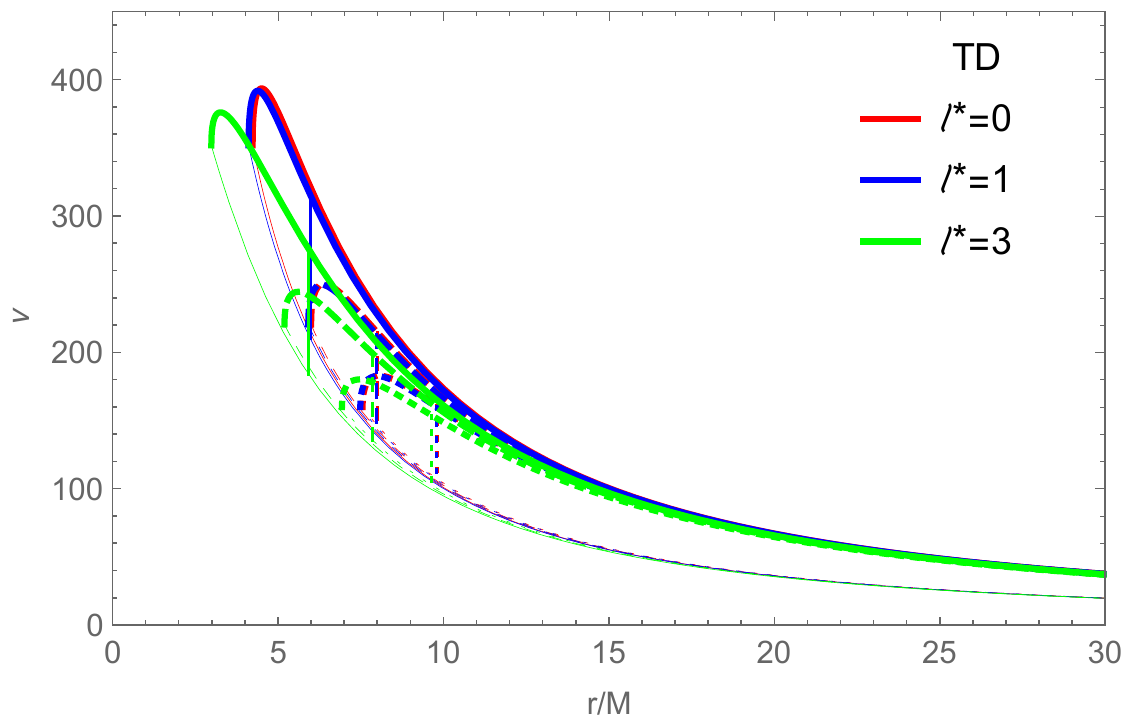}
\includegraphics[width=5.5cm]{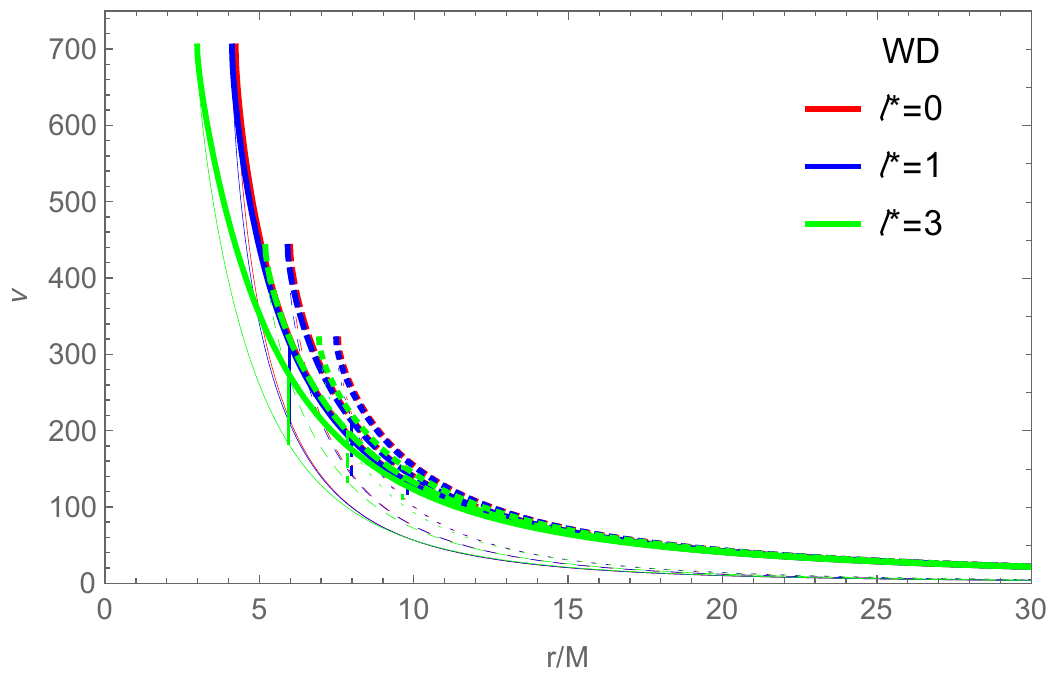}
  \caption{In the cases of the standard RP and its variants, TD and WD model the particle undergoes epicycle motion around three types of RSV celestial bodies with spin parameters $\alpha^*=$ 0.5 (solid line), $\alpha^*=0$ (dashed line), and $\alpha^*=-0.5$ (dotted line), with the values of the twin peaks frequency $v_u$ (thick line) and $v_l$ (thin line) varying with the radial coordinate $r/M$. The vertical lines represent the positions of the frequency ratio $v_u: v_l=3: 2$.}\label{fig:5}
\end{figure}

In the following, we use $r=r_{\text {ISCO }}$ as the starting point for the radial coordinate in Figures \ref{fig:4} and \ref{fig:5}, taking the spin parameter $\alpha^*$ as 0.5 (solid line), 0 (dashed line), and -0.5 (dotted dashed line) as examples. We plot the twin peak frequency values of particles undergoing oscillatory motion around different types of central celestial bodies (Kerr black hole $l^*=0$, regular black hole $l^*=1$, traversable wormhole $l^*=3$) in the oscillatory motion of ER $\left(\mathrm{ER}_0, \mathrm{ER}_1, \mathrm{ER}_2, \mathrm{ER}_3, \mathrm{ER}_4, \mathrm{ER}_5\right)$, RP $\left(\mathrm{RP}_0, \mathrm{RP}_1\, \mathrm{RP}_2\right)$, TD, and WD models. The points at which the $3: 2$ resonance occurs under different conditions are also marked. The thick and thin lines represent high and low frequencies respectively. A central object mass of $M=10 M_{\odot}$ is assumed in the plot.

The RP model, proposed by Stella and Vietri \cite{82}, is a kinematic model, in which QPOs are generated by the local motion of plasma within the accretion disk surrounding the central celestial body \cite{107}. For the standard RP ($\mathrm{RP}_0$) model, the high frequency in twin peaks HFQPOs is represented by the orbital frequency of the test particle $\left(v_u=v_0\right)$, while the low frequency is represented by a linear combination of the particle's orbital frequency and the epicyclic motion radial frequency $\left(v_l=v_0-v_r\right)$. Additionally, there are two variants of the RP model: $\mathrm{RP}_1$ and $\mathrm{RP}_2$ [101]. The high and low frequencies of twin peaks for both variants are denoted as $v_u=v_\theta, v_l=v_0-v_r$ and $v_u=v_0, v_l=v_\theta-v_r$.

The TD model \cite{84,85,86,107,108} is also a kinematic model. This model suggests that HFQPO may be caused by a tidal stretching of the plasma around the central object. Under the effect of tidal stretching, the plasma forms ring-like structures along the orbit, leading to the modulation of the observed radiation flux in the black hole power spectrum\cite{107}. The twin peaks frequency model is $v_u=v_0+v_r, v_l=v_0$. Finally, in the WD model, the HFQPO phenomenon occurs due to the oscillations of a warped thin disk. This model explains HFQPOs through a nonlinear resonance mechanism. These resonances primarily occur between the relativistic disk, deformed by warp effects, and various disk oscillation modes \cite{87,107,WD-1}. The origin of these resonances lies in the non-monotonic distribution of the radial epicyclic frequency with respect to
$r$ \cite{87}. The upper and lower frequencies of this model are defined as $v_u=2 v_0-v_r$ and $v_l=$ $2\left(v_0-v_r\right)$ \cite{87,108}.

From Figures \ref{fig:4}, \ref{fig:5}, we can observe that in the ER $\left(\mathrm{ER}_0, \mathrm{ER}_1, \mathrm{ER}_2, \mathrm{ER}_3, \mathrm{ER}_4, \mathrm{ER}_5\right)$, RP $\left(\mathrm{RP}_0, \mathrm{RP}_1, \mathrm{RP}_2\right)$, TD, and WD models, particles undergoing oscillatory motion around the RSV celestial bodies of positive rotation $(\alpha^*=0.5)$ exhibit a twin peaks frequency value, which is greater than that in the stationary $(\alpha^*=0)$ condition and greater than that in the counter-spin $(\alpha^*=-0.5)$ case. Concurrently, through the images, one can perceive the location of the twin peaks frequency ratio $v_u: v_l=$ $3: 2$, which represents the position of resonance within the accretion disc surrounding the central celestial body.

Since resonance can significantly change the dynamics of test particles around the central object, finding the location of resonance is important for studying the motion of accreting fluid around the central object and exploring the physical mechanism of HFQPOs \cite{78}. In all of the HFQPOs models we discussed, the location of the 3:2 resonance occurring in the three types of central objects moves away from the radial coordinate center as the $\alpha^*$ value decreases. Furthermore, we observed the deviation of the resonance position of RSV regular black holes and traversable wormholes in different HFQPO models relative to Kerr black holes. Only in the $\mathrm{ER}_0$ and $\mathrm{ER}_1$ models, the resonance positions of regular black holes and traversable wormholes are further away from the radial coordinate center compared to the Kerr black hole. However, in the other nine HFQPO models studied in this paper, the resonance radii of regular black holes and traversable wormholes are all smaller than the resonance radius of the Kerr black hole.

\section{$\text{Constraints on the parameters of RSV spacetime using HFQPO observational data}$}
\subsection{{Data fitting and results}}
In recent years, testing gravitational theories using astronomical data has become a significant topic in relativistic astrophysics. The data used include HFQPOs observed in microquasars \cite{71,72,73,74,75,76,77,78}, the black hole shadow of Sgr A* observed through telescopes \cite{2,3}, and the hotspot data from three flares observed via gravitational instruments \cite{107}, among others. In these observations, microquasars are considered as candidates for stellar-mass black holes. However, obtaining precise information about these black holes is difficult. Therefore, observational HFQPO data can be employed to test gravitational theories (black hole or wormhole solutions) \cite{108}.

In the following, we explore the limits on RSV spacetime according to the observational HFQPO data from three microquasars (detailed in Table \ref{table2}) \cite{109,110,111,112,113}, labeled as GRO 1655-40 ($k_1$), XTE 1550-564 ($k_2$), and GRS 1915+105 ($k_3$). The specific data include the high and low frequencies of HFQPOs' double peaks and the masses $M / M_{\odot}$ of the three microquasar sources determined via the optical/NIR photometry, where $M_{\odot}$ represents the solar mass \cite{114,115}. Assuming that these microquasars can be described by the RSV spacetime, we use the observational data to constrain the spin parameters $\alpha_k^*$ $(k=1,2,3)$ of the three microquasars and RSV regularization parameter $l^*$. Additionally, we test the eleven common HFQPO models introduced in the previous section aiming to search for the HFQPO models that matches the observational data, which could help us explore the potential physical mechanisms that generate the HFQPOs.

\begin{table}[ht]
\begin{center}
\begin{tabular}{|c|c|c|c|c|c|c|}
\hline Source & $v_{1 u}[Hz]$ & $v_{1 l}[Hz]$ & $v_{1 c}[Hz]$ & $v_{2 u}[Hz]$ & $v_{2 c}[Hz]$ & $M / M_{\odot}$ \\
\hline GRO 1655-40 $\left(k_1\right)$ & $441 \pm 2$ & $298 \pm 4$ & $17.3 \pm 0.1$ & $451 \pm 5$ & $18.3 \pm 0.1$ & $5.4 \pm 0.3$ \\
\hline XTE 1550-564 $\left(k_2\right)$ & $276 \pm 3$ & $184 \pm 5$ & --- & --- & --- & $9.1 \pm 0.61$ \\
\hline GRS 1915+10 $\left(k_3\right)$ & $168 \pm 3$ & $113 \pm 5$ & --- & --- & --- & $12.4_{-1.8}^{+2.0}$ \\
\hline
\end{tabular}
\end{center}
\caption{\label{table2}
   The observed HFQPO data from three microquasars, where $v_{c} =v_0 - v_\theta $ \cite{109,110,111,112,113,114,115}.}
\end{table}

In Table \ref{table2}, we select four sets of observational data, two sets of observational data for GRO 1655-40 \cite{111}. Both sets of GRO 1655-40 data include the nodal precession frequency $(v_{c} =v_0 - v_\theta) $. For the microquasars XTE 1550-564 and GRS 1915+10, we select the widely recognized sets of observational data available so far \cite{112,113}. However, no studies have yet provided observational reports on the nodal precession frequency $\left(v_c\right)$ for these two microquasars (in the corresponding positions in the Table \ref{table2}, we use "$-$" to mark).
Consider that the four sets of HFQPO data in Table \ref{table2} are generated at distinct circular orbits with radii $r_1, r_1^{\prime}, r_2$ and $r_3$.
As indicated by equations (\ref{9}), (\ref{17}), and (\ref{18}), besides the resonance locations, the theoretical model includes four free parameters: the regularization parameter $l^*$, and the spin parameters $\alpha_k^*$ of the three microquasars. To determine the values of the model parameters, we perform $\chi^2$ analyses using equation (\ref{20}) for $\mathrm{E R}_i$ $(i=0, 1, 2, 3, 4, 5)$, $\mathrm{T D}$, $\mathrm{W D}$ models and equation (\ref{21}) for $\mathrm{R P}_j$ $(j=0, 1, 2)$ models

\begin{equation}
\chi_1^2=\frac{\left\{v_{2 u, 1}-v_{2 u}\left(l, \alpha_1, r_1^{\prime}\right)\right\}^2}{\sigma_{v_{2 u, 1}}^2}+\sum_{k=1}^3 \frac{\left\{v_{1 u, k}-v_{1 u}\left(l, \alpha_k, r_k\right)\right\}^2}{\sigma_{v_{1 u, k}}^2}+\sum_{k=1}^3 \frac{\left\{v_{1 l, k}-v_{1 l}\left(l, \alpha_k, r_k\right)\right\}^2}{\sigma_{v_{1 l, k}}^2}, \label{20}
\end{equation}
\begin{equation}
\begin{gathered}
\chi_2^2=\frac{\left\{v_{1 c, 1}-v_{1c}\left(l, \alpha_1, r_1\right)\right\}^2}{\sigma_{v_{1 c,1}}^2}+\frac{\left\{v_{2 c, 1}-v_{2c}\left(l, \alpha_1, r_1^{\prime}\right)\right\}^2}{\sigma_{v_{2 c, 1}}^2}+\frac{\left\{v_{2 u, 1}-v_{2 u}\left(l, \alpha_1, r_1^{\prime}\right)\right\}^2}{\sigma_{v_{2 u, 1}}^2} \\
+\sum_{k=1}^3 \frac{\left\{v_{1 u, k}-v_{1 u}\left(l, \alpha_k, r_k\right)\right\}^2}{\sigma_{v_{1 u, k}}^2}+\sum_{k=1}^3 \frac{\left\{v_{1 l, k}-v_{1 l}\left(l, \alpha_k, r_k\right)\right\}^2}{\sigma_{v_{1 l, k}}^2}.
\end{gathered} \label{21}
\end{equation}

\begin{figure}[ht]
\includegraphics[width=5.5cm]{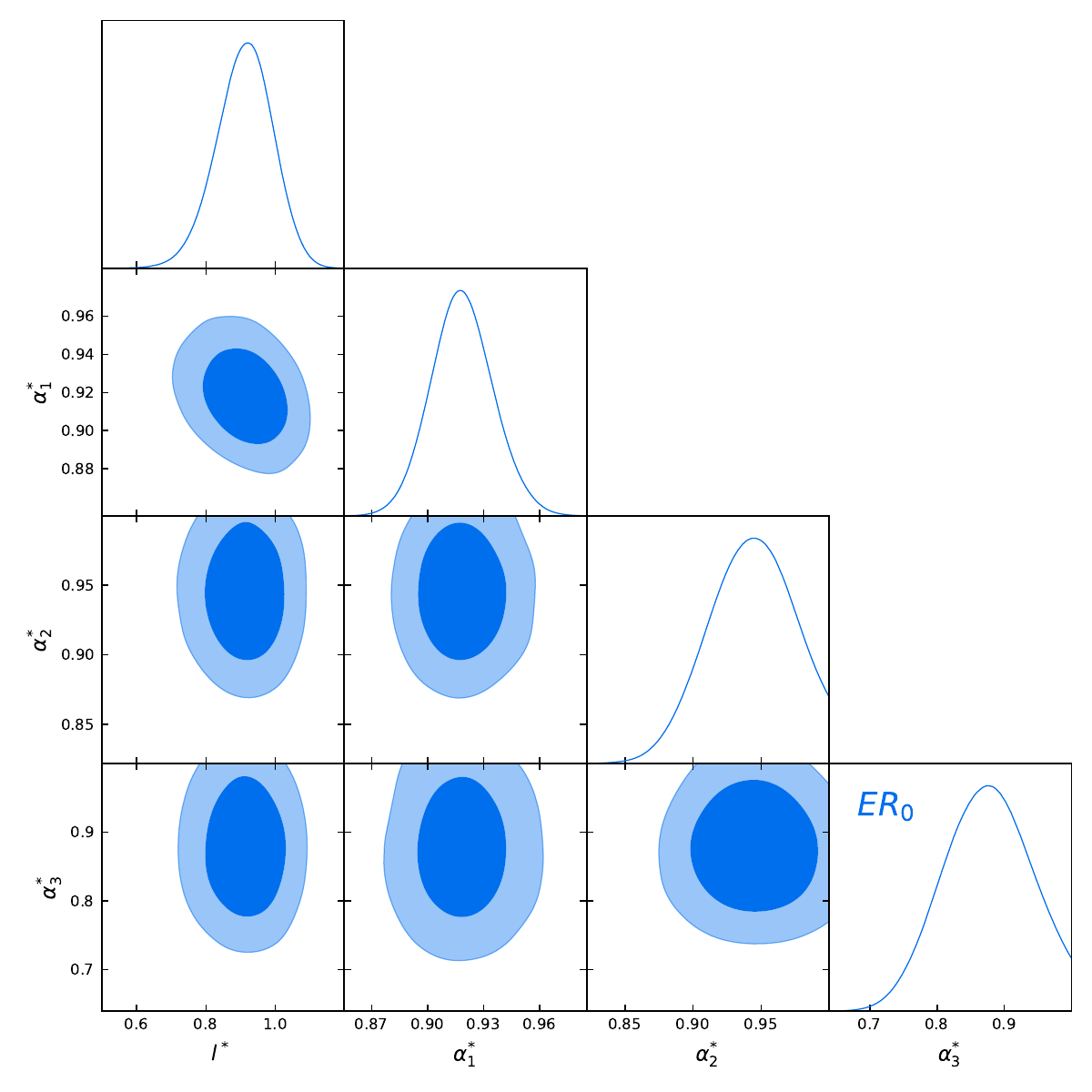}
\includegraphics[width=5.5cm]{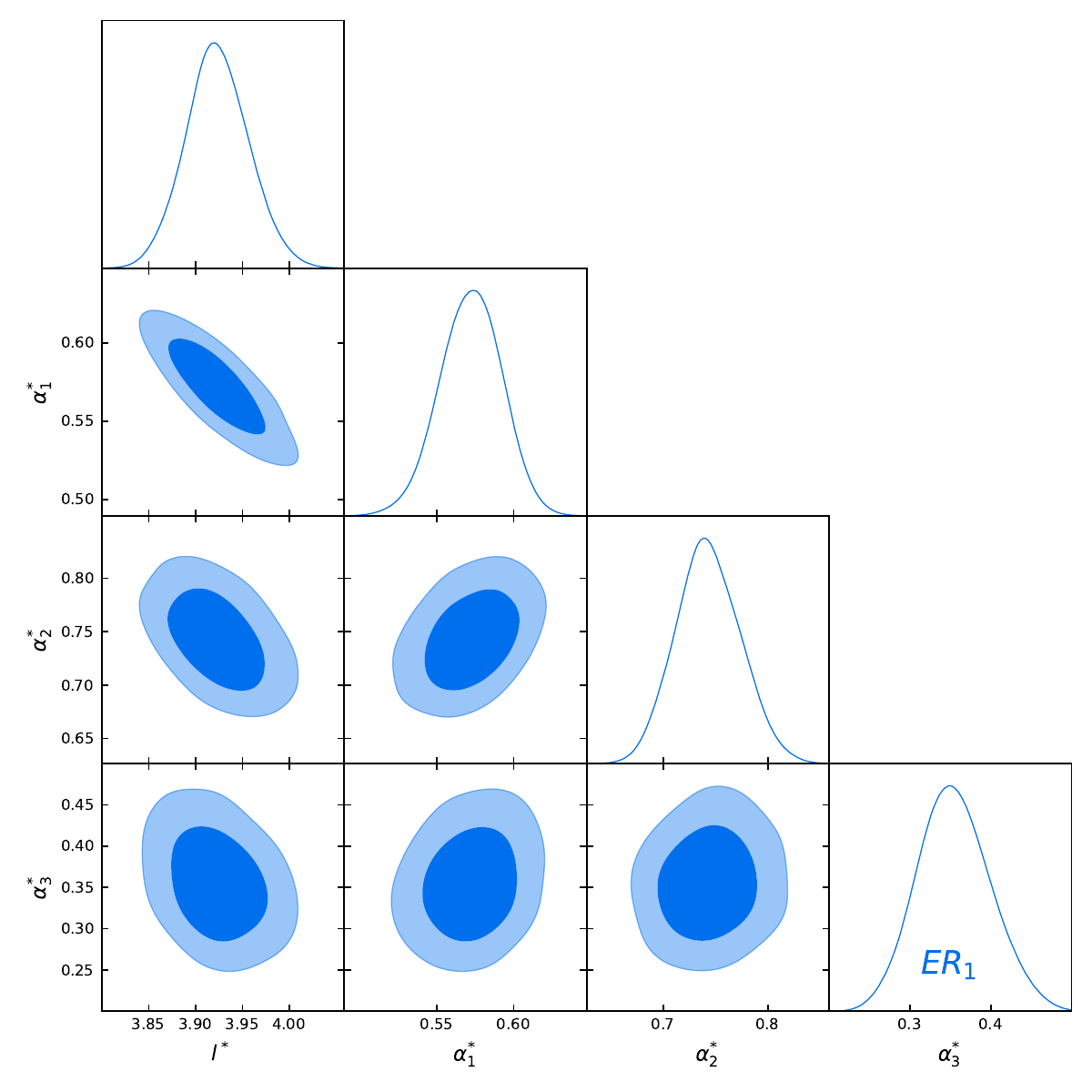}
\includegraphics[width=5.5cm]{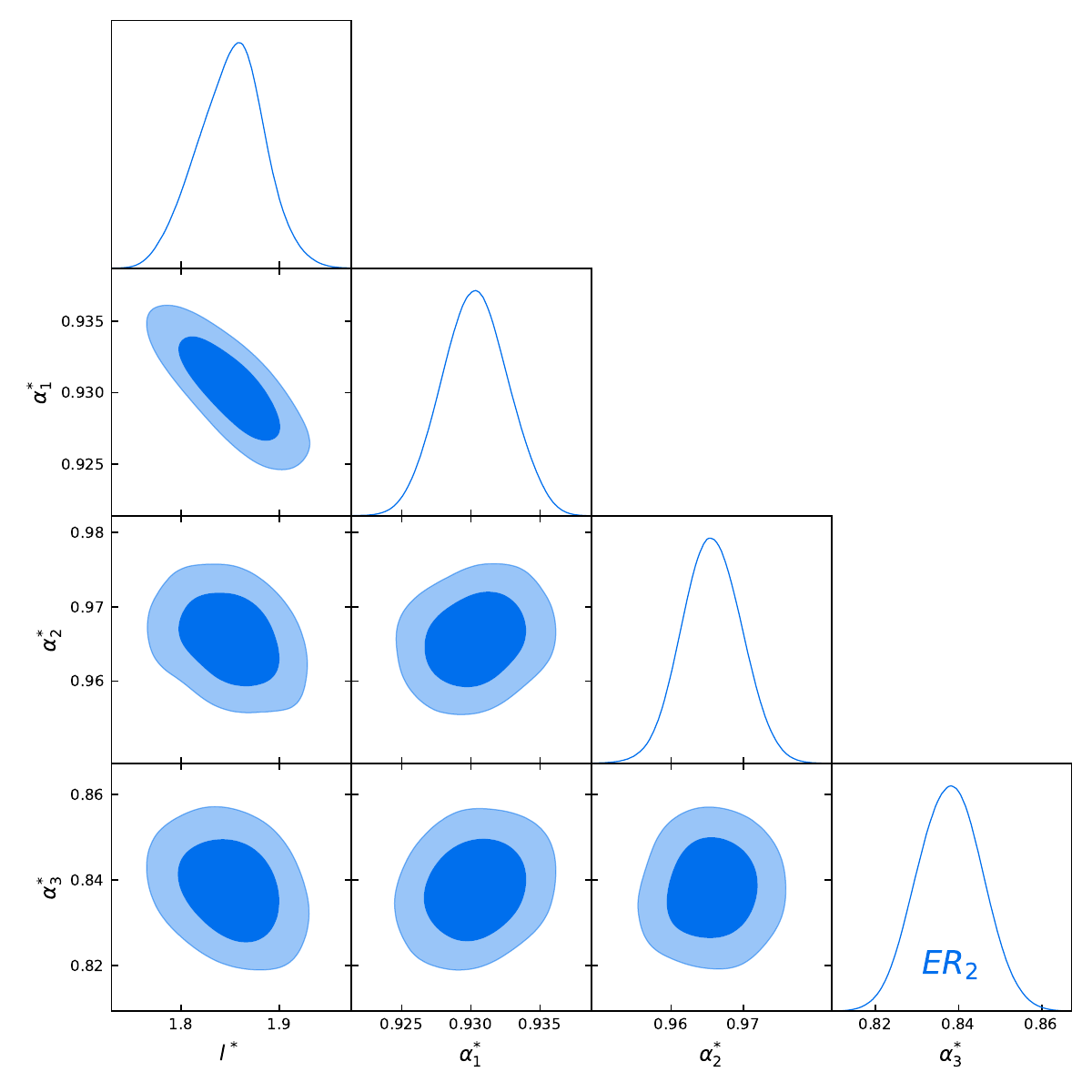}
\includegraphics[width=5.5cm]{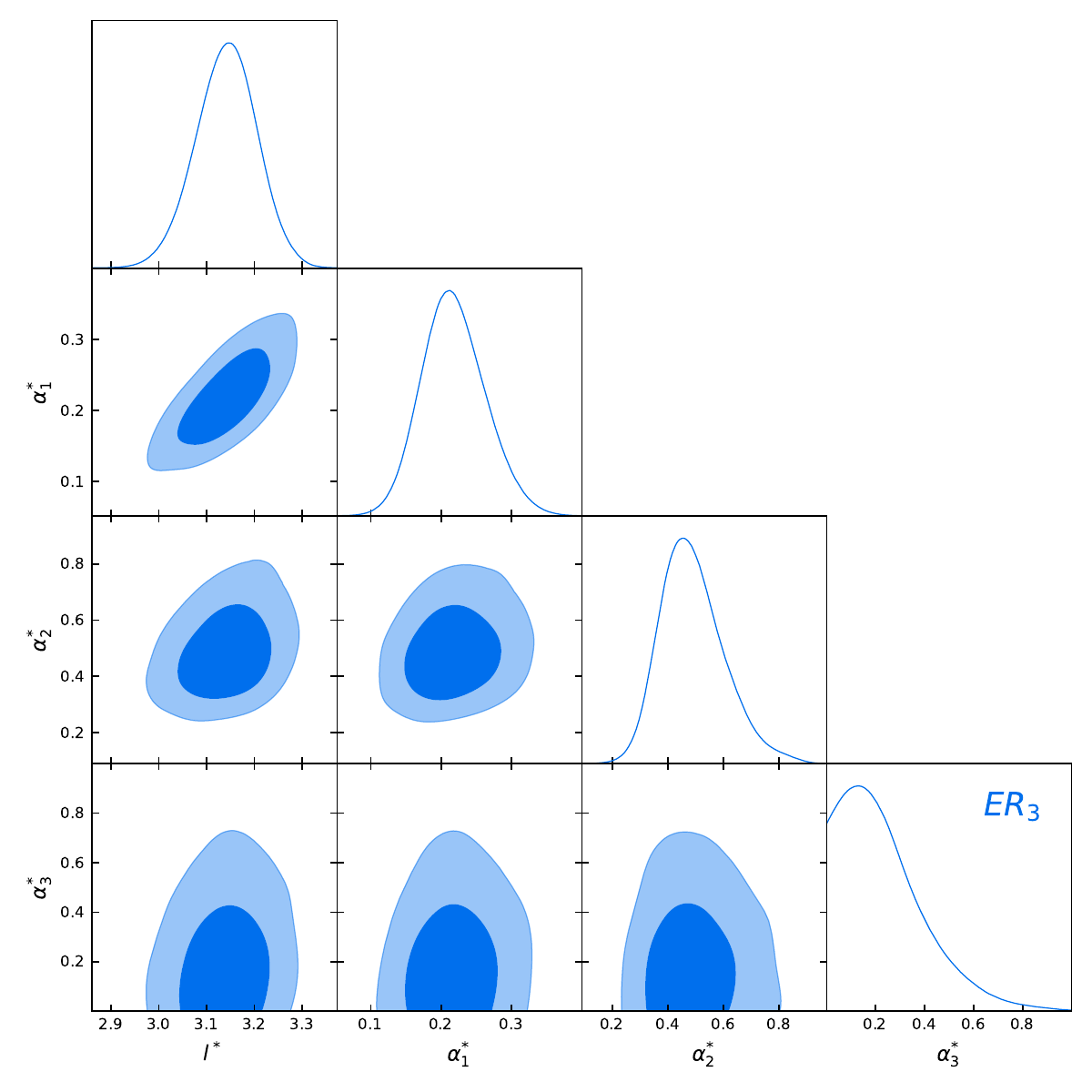}
\includegraphics[width=5.5cm]{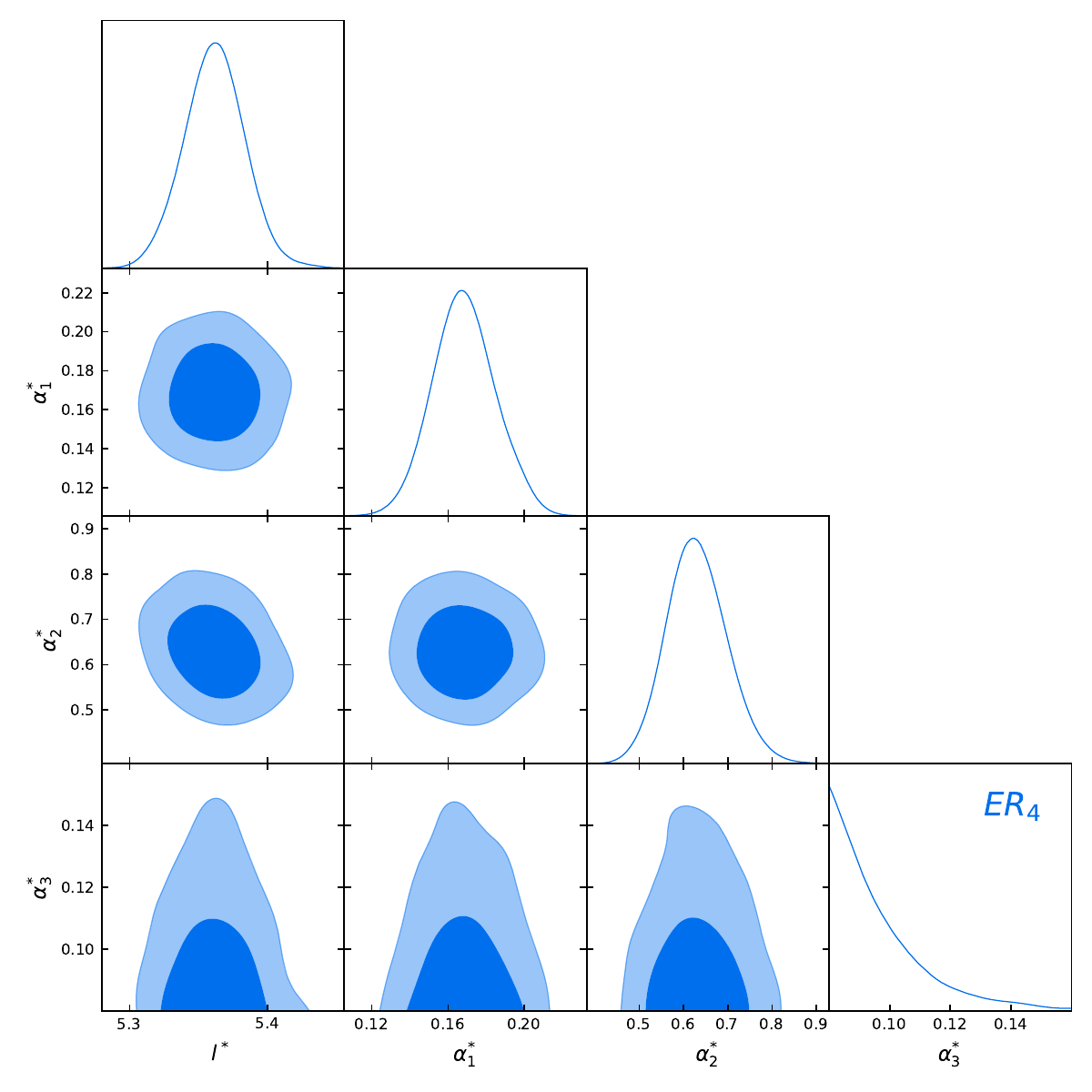}
\includegraphics[width=5.5cm]{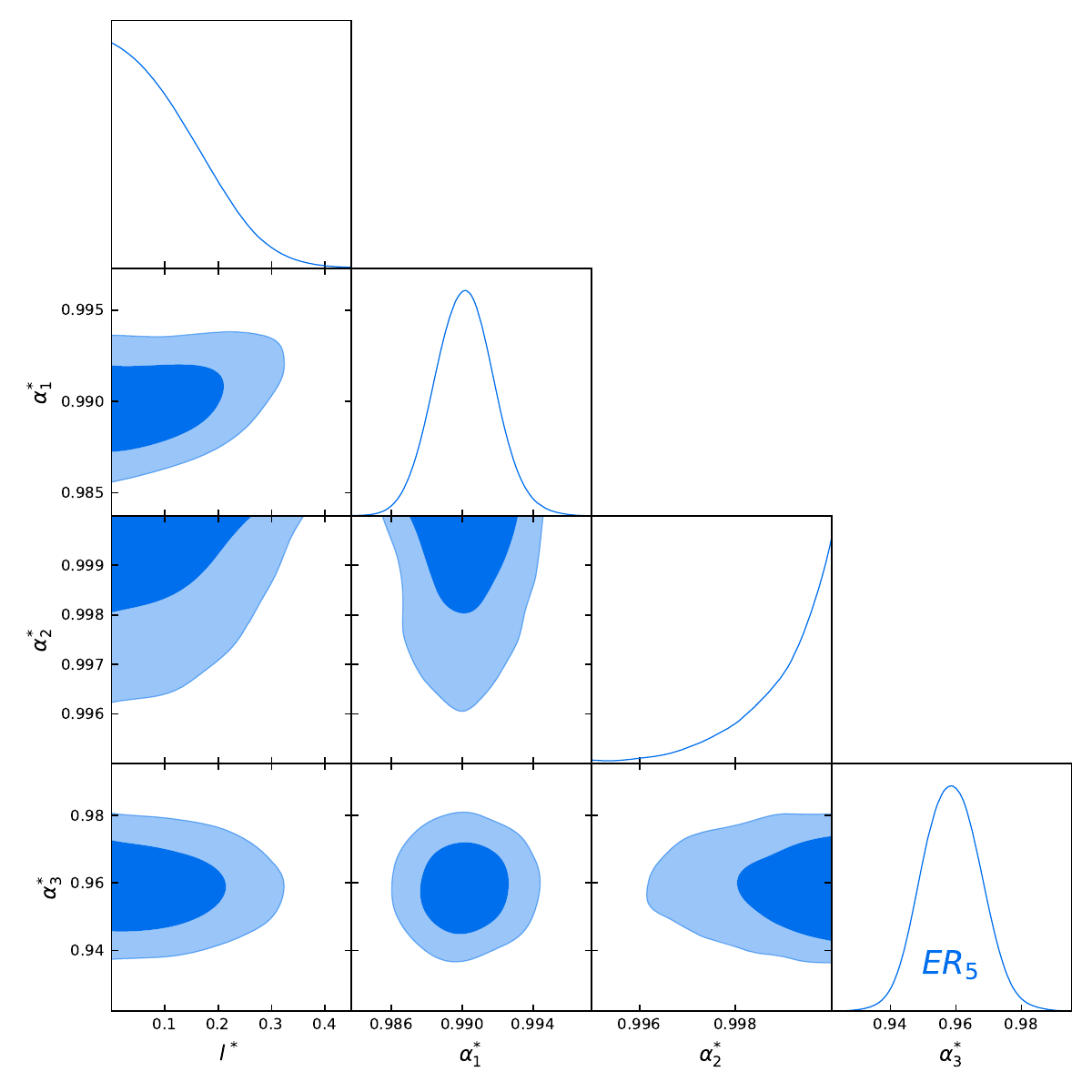}
  \caption{In RSV sapcetime, the $68 \%$ and $95 \%$ confidence regions of the regularization parameter $l^*$ and the spin parameters $\alpha_k^*$ of the three microquasars under $\mathrm{ER}_0 - \mathrm{ER}_5$ models.}\label{fig:6}
\end{figure}

\begin{figure}[ht]
\includegraphics[width=5.5cm]{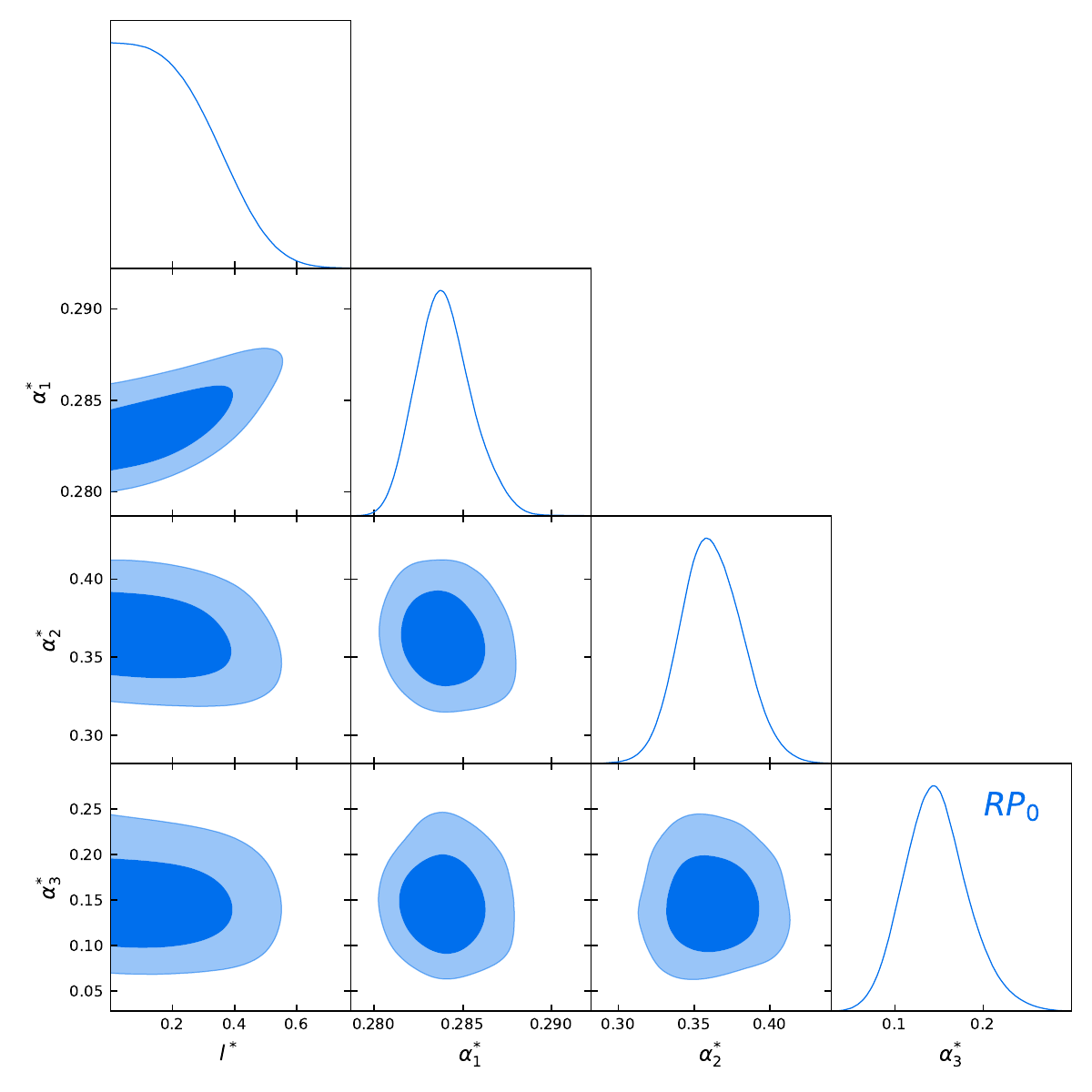}
\includegraphics[width=5.5cm]{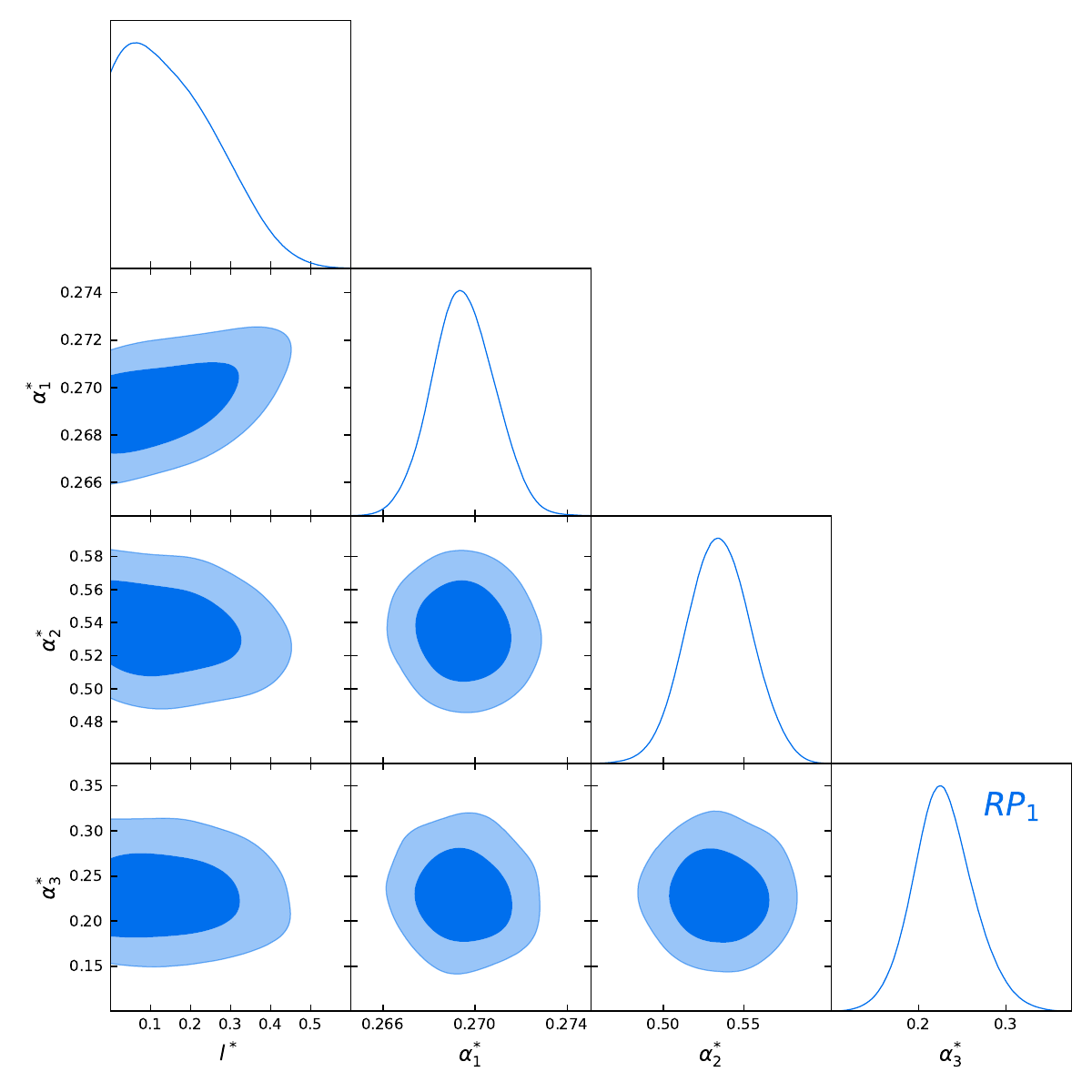}
\includegraphics[width=5.5cm]{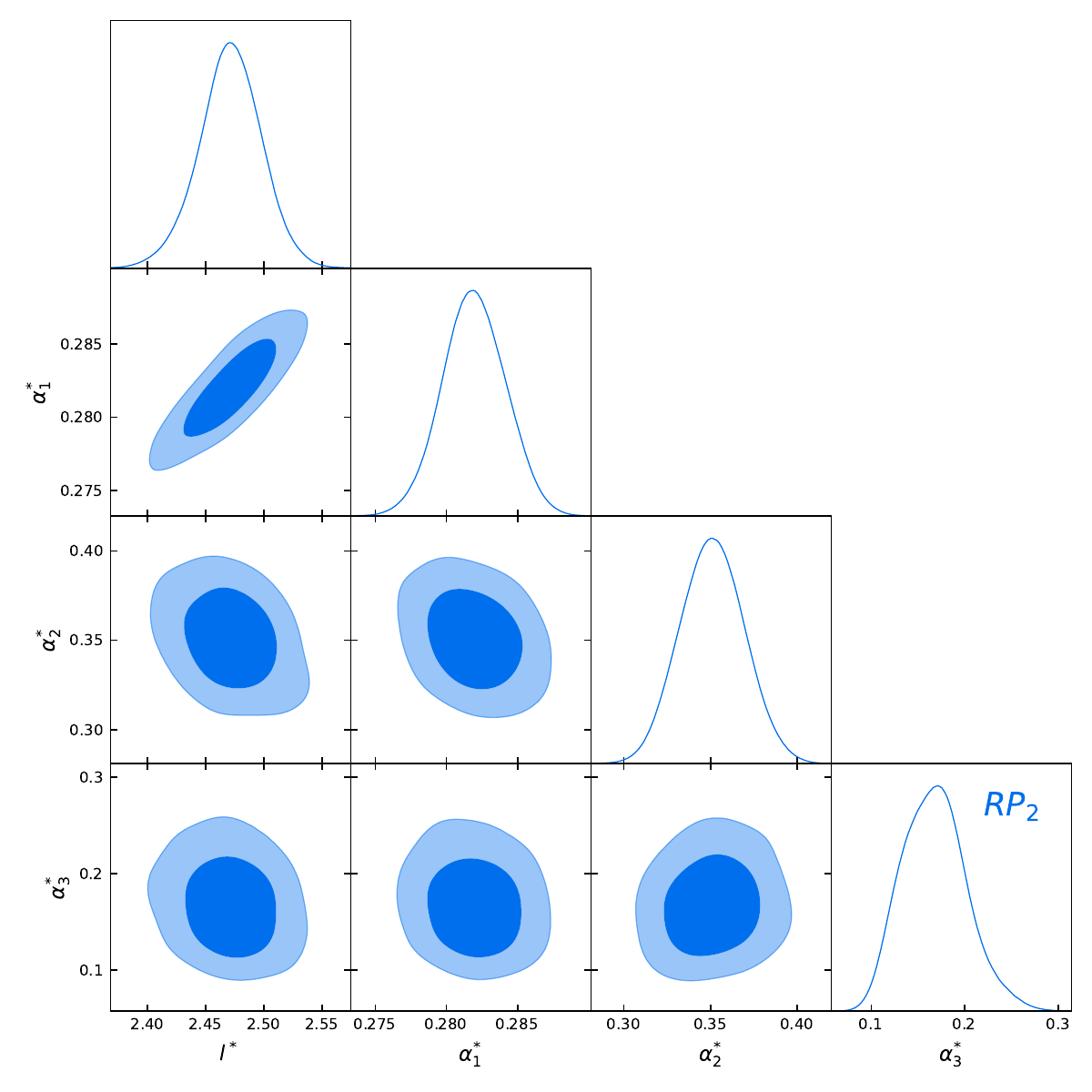}
\includegraphics[width=5.5cm]{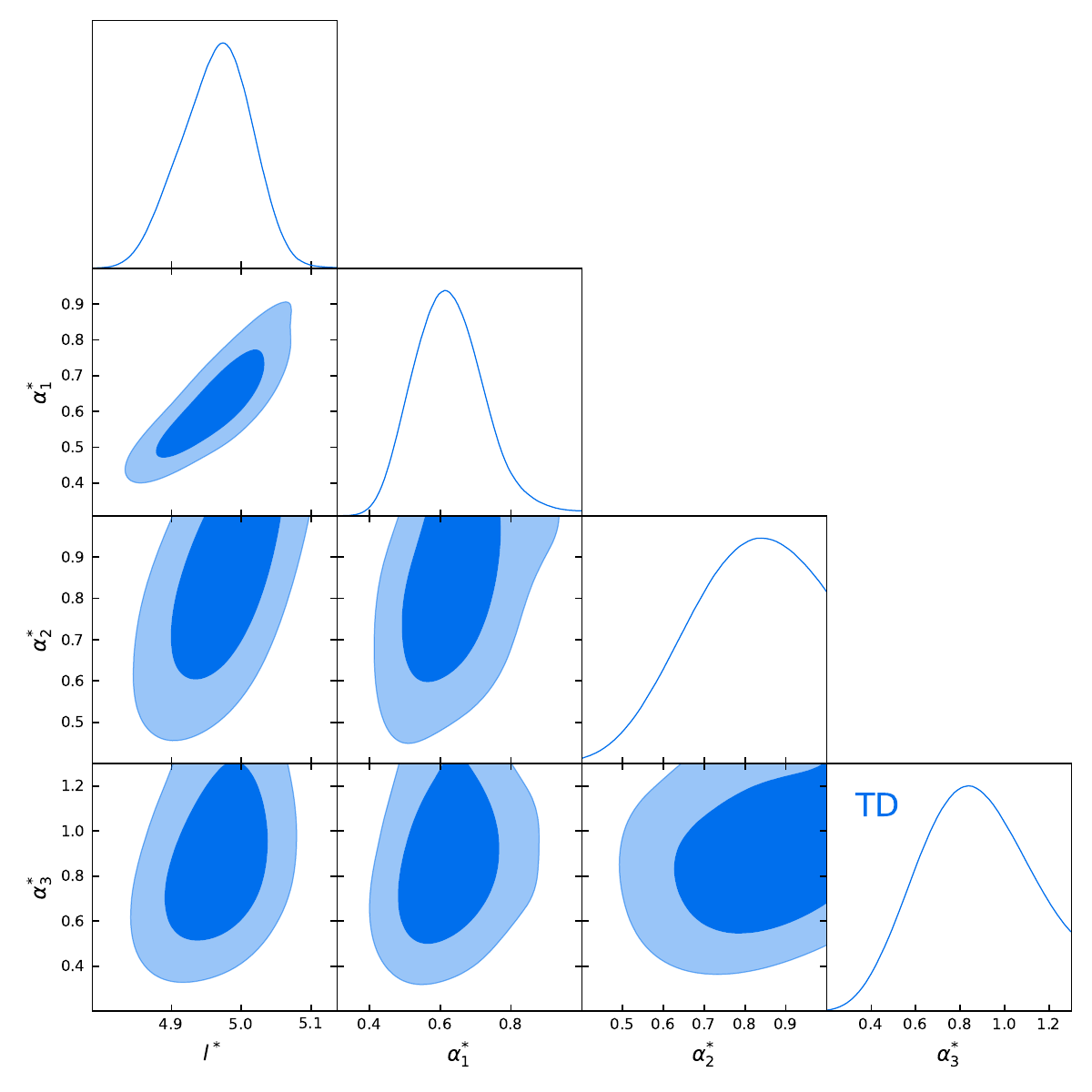}
\includegraphics[width=5.5cm]{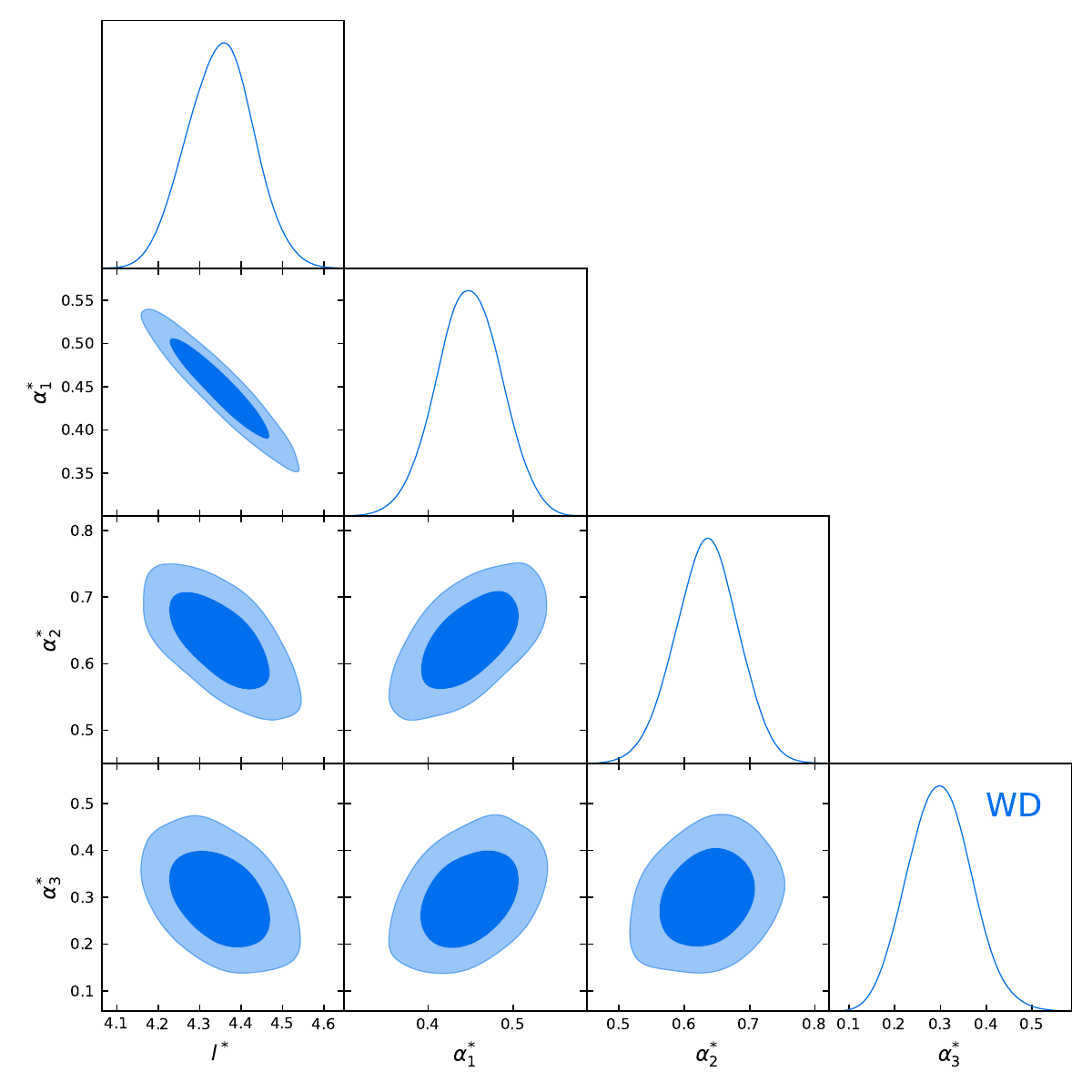}
  \caption{In RSV sapcetime, the $68 \%$ and $95 \%$ confidence regions of the regularization parameter $l^*$ and the spin parameters $\alpha_k^*$ of the three microquasars under $\mathrm{RP}_0$, $\mathrm{RP}_1$, $\mathrm{RP}_2$, $\mathrm{TD}$ and $\mathrm{WD}$ models.}
    \label{fig:7}
\end{figure}

Based on the constraints from the HFQPO data, Table \ref{table3} summarizes the best-fit values for the regularization parameter $l^*$ and the spin parameters $\alpha_k^*$ of the three microquasars within the $68 \%$ confidence level (CL), as well as the value of $\chi_{\text {min }}^2$. In addition, Table \ref{table4} presents the best-fit values for the orbital radii (i.e., resonance locations) associated with the four sets of HFQPO data. Finally, Figures \ref{fig:6} and \ref{fig:7} illustrate the RSV parameter plots for the regularization parameter $l^*$ and the spin parameters $\alpha_k^*$ of the three microquasars within the $68 \%$ and $95 \%$ CL under different HFQPO models.

\begin{table}[ht]
\begin{center}
\begin{tabular}{|c|c|c|c|c|c|c|}
\hline & $E R_0$ & $E R_1$ & $E R_2$ & $E R_3$ & $E R_4$ & $E R_5$ \\
\hline $l^*$ & $0.908_{-0.073}^{+0.086}$ & $3.927 \pm 0.033$ & $1.850 \pm 0.036$ & $3.110 \pm 0.070$ & $5.360 \pm 0.021$ & $<0.138$ \\
\hline $\alpha_1^*$ & $0.918 \pm 0.016$ & $0.570 \pm 0.019$ & $0.930 \pm 0.003$ & $0.220_{-0.051}^{+0.041}$ & $0.167 \pm 0.016$ & $0.990_{-0.018}^{+0.002}$ \\
\hline $\alpha_2^*$ & $0.942_{-0.027}^{+0.035}$ & $0.744_{-0.037}^{+0.030}$ & $0.966 \pm 0.004$ & $0.490_{-0.140}^{+0.100}$ & $0.640 \pm 0.067$ & $<0.999$ \\
\hline $\alpha_3^*$ & $0.870 \pm 0.062$ & $0.349 \pm 0.045$ & $0.838_{-0.008}^{+0.087}$ & $<0.299$ & $<0.100$ & $0.959 \pm 0.009$ \\
\hline $\chi_{\text {min }}^2$ & 0.301 & 0.152 & 0.183 & 7.644 & 15.846 & 9.656 \\
\hline & $R P_0$ & $R P_1$ & $R P_2$ & $TD$ & $WD$ & \\
\hline $l^*$ & $<0.314$ & $<0.247$ & $2.496_{-0.027}^{+0.024}$ & $4.964 \pm 0.046$ & $4.358 \pm 0.083$ & \\
\hline $\alpha_1^*$ & $0.284_{-0.002}^{+0.001}$ & $0.270 \pm 0.002$ & $0.282 \pm 0.002$ & $0.625_{-0.098}^{+0.081}$ & $0.445 \pm 0.041$ & \\
\hline $\alpha_2^*$ & $0.358 \pm 0.022$ & $0.536_{-0.022}^{+0.018}$ & $0.352_{-0.018}^{+0.016}$ & $0.780_{-0.100}^{+0.018}$ & $0.629_{-0.054}^{+0.063}$ & \\
\hline $\alpha_3^*$ & $0.162 \pm 0.041$ & $0.231 \pm 0.035$ & $0.166_{-0.040}^{+0.031}$ & $0.840 \pm 0.160$ & $0.266 \pm 0.076$ & \\
\hline $\chi_{\text {min }}^2$ & 1.195 & 65.632 & 0.387 & 0.097 & 0.158 & \\
\hline
\end{tabular}
\end{center}
\caption{\label{table3}
   In RSV spacetime, the best-fit values of the regularization parameter $l^*$ and the spin parameters $\alpha_k^*$ of the three microquasars within the $68 \%$ $\mathrm{CL}$, as well as the minimum $\chi^2$ value, for different resonance models.}
\end{table}

\begin{table}[ht]
\begin{center}
\begin{tabular}{|c|c|c|c|c|c|c|c|c|c|c|c|}
\hline & $E R_0$ & $E R_1$ & $E R_2$ & $E R_3$ & $E R_4$ & $E R_5$ & $R P_0$ & $R P_1$ & $R P_2$ & $T D$ & $W D$ \\
\hline$r_1 / M$ & 4.865 & 3.440 & 2.370 & 6.646 & 3.280 & 2.952 & 5.628 & 5.555 & 5.053 & 5.526 & 5.969 \\
\hline$r_1^{\prime} / M$ & 4.761 & 3.399 & 2.356 & 6.591 & 3.050 & 2.611 & 5.526 & 5.423 & 4.933 & 5.309 & 5.865 \\
\hline$r_2 / M$ & 4.638 & 2.883 & 2.111 & 6.311 & 1.984 & 2.782 & 5.412 & 5.031 & 4.808 & 5.310 & 5.568 \\
\hline$r_3 / M$ & 5.469 & 4.470 & 3.057 & 7.552 & 4.612 & 3.437 & 6.197 & 6.034 & 5.662 & 5.367 & 6.795 \\
\hline
\end{tabular}
\end{center}
\caption{\label{table4}
   The best-fit values of the circular orbit radii associated with the four sets of HFQPOs data under different resonance models, i.e., the orbital radii that produce the HFQPOs (resonance positions).}
\end{table}

\subsection{{Model selection}}

We can see that the constraints on RSV spacetime in the previous chapter are model-dependent, with the results being significantly influenced by assumptions about the mechanism responsible for generating HFQPOs. The mechanism of HFQPOs, however, remains unclear at present. Still, it indicates that tests based on HFQPOs could have the huge potential, since the clearer understanding of their origin would allow for more precise testing of the spacetimes under investigation.
From section 6.1, we can see that for $\mathrm{E R}_i$ $(i=0, 1, 2, 3, 4, 5)$, $\mathrm{T D}$ and $\mathrm{W D}$ models, the used data points to constrain parameters are completely identical. However, the $\mathrm{R P}_j$ $(j=0, 1, 2)$ models include two additional data points compared to the aforementioned eight models. Therefore, in this section, we first apply the limit results to compute the $AIC$ to assess the fit of the first eight HFQPO models (including $\mathrm{E R}_i$ $(i=0, 1, 2, 3, 4, 5)$, $\mathrm{T D}$, $\mathrm{W D}$) to the HFQPO observational data, thereby identifying which of these models are more supported by the observational data. Subsequently,
we perform a comparison between the best model selected by the $A I C$ and the $\mathrm{R P}_j(j=0,1,2)$ models by calculating the Bayes factors, in order to further assess them using the observational data.

The $AIC$ was initially introduced to cosmology by Liddle \cite{116}, and later extended to other fields of research \cite{117,118}. Its definition is given as \cite{119}
\begin{equation}
A I C=-2 \ln \mathcal{L}(\Theta\mid D)_{\max }+2 K. \label{22}
\end{equation}
In this equation, $D$ is the data, $K$ represents the number of estimable parameters $\Theta$ in the model, and $\mathcal{L}_{\max }$ denotes the maximum likelihood achieved with the best-fit parameters $\Theta$. The term $-2 \ln \mathcal{L}(\Theta\mid D)_{\max }$ in Eq.(\ref{22}) is referred to as $\chi^2$, which evaluates the quality of the model's fit, while the term $2 K$ in Eq.(\ref{22}) accounts for the model's complexity \cite{120,121,122,123,124}.

The $AIC$ value of an individual model has no inherent meaning; rather, the relative values among different models are what carry practical significance. As such, the model with the lowest $AIC$ value is considered the optimal model, expressed as $A I C_{\min }=\min \left\{A I C_i, i=1, \ldots, N\right\}$, where $A I C_i$ represents a set of alternative candidate models. By calculating the likelihood of model $\mathcal{L}\left(\bar{M}_i \mid\right.$ D $) \propto$ $\exp \left(-\triangle_i / 2\right)$, the relative evidential strength of each model can be determined, with $\triangle_i=A I C_i-A I C_{\text {min }}$ across the entire range of candidate models. The criteria for model selection are as follows:

A: $0 < \triangle_i \leq 2$, model $i$ receives nearly the same level of support from the data as the best model;

B: $4<\triangle_i \leq 10$, model $i$ is significantly less supported;

C: $\triangle_i>10$, model $i$ is essentially irrelevant to the observational data.

Based on the calculation results, the best model is $\mathrm{TD}$, with an $AIC$ value of 8.097. Taking it as a reference, we can compare the models in Table \ref{table3} through the $AIC$ difference $\triangle_i$. Considering that HFQPO models ($\mathrm{E R}_i$ $(i=0, 1, 2, 3, 4, 5)$, $\mathrm{T D}$, $\mathrm{W D}$) have the same number of parameters, the $\triangle_i$ can also be calculated equally by using the $\triangle \chi_{\min }^2$ value. It is important to note that $AIC$ model selection provides quantitative insights into the "strength of evidence," rather than merely identifying the best model. Below, we classify the HFQPO models within different $\triangle_i$ ranges (see Table \ref{table6}).
\begin{table}[ht]
\begin{center}
\begin{tabular}{|l|l|c|c|c|}
\hline & $\triangle_i=0$ & $0<\triangle_i \leqslant 2$ & $4<\triangle_i \leqslant 10$ & $\triangle_i>10$ \\
\hline HFQPO models & $T D$ & $E R_0, E R_1, E R_2, WD$ & $E R_3, E R_5$ & $E R_4$\\
\hline
\end{tabular}
\end{center}
\caption{\label{table6}
    Relative to the best model $\mathrm{TD}$, the $\triangle_i$ intervals for other HFQPO models.}
\end{table}

From Table \ref{table6}, it can be seen that the differences in support for different models by the HFQPO data are significant. The results show that $\mathrm{ER}_0$, $\mathrm{ER}_1$, $\mathrm{ER}_2$,  and $\mathrm{WD}$ models have the same support by observational data ($0 < \triangle_i \leq 2$) as the best model $\mathrm{TD}$, whereas the $\mathrm{ER}_3, \mathrm{ER}_5$ models receive the moderate levels of evildential support. Furthermore, $\mathrm{ER}_4$ models are no substantial evidence.

Next, we compare the best model ($\mathrm{TD}$) selected using the $AIC$ method with the $R P_j(j=0,1,2)$ models through Bayes factors to further evaluate them by the observational data. In the framework of Bayesian inference, we focus on the posterior probability $P(\Theta \mid D, \bar{M})$, and $\bar{M}$ is the model. The posterior is calculated using the following equation \cite{Bayes-1, Bayes-2}

\begin{equation}
P(\Theta \mid D, \bar{M})=\frac{P(D \mid \Theta, \bar{M}) \pi(\Theta \mid \bar{M})}{P(D \mid \bar{M})}, \label{30}
\end{equation}
where $\pi(\Theta)$ denotes the prior distribution of $\Theta$, and $P(D \mid \bar{M})$ is expressed as
\begin{equation}
P(D \mid \bar{M})=\int P(D \mid \Theta, \bar{M}) \pi(\Theta) d \Theta. \label{31}
\end{equation}
Similarly, for the model $\bar{M}$, we have
\begin{equation}
P(\bar{M} \mid D) \propto P(D \mid \bar{M}) \pi(\bar{M}) \propto Z_{\bar{M}} \pi_{\bar{M}}, \label{32}
\end{equation}
where $Z_{\bar{M}}=P(D \mid \bar{M})$ is the model evidence, and $\pi_{\bar{M}}=\pi(\bar{M})$ is the prior probability of the model. To compare models, the Bayes factor is defined as
\begin{equation}
R=\frac{Z_{\bar{M}_1} \pi_{\bar{M_1}}}{Z_{\bar{M_2}} \pi_{\bar{M}_2}}. \label{33}
\end{equation}
If $R>1$, it indicates that $\bar{M}_1$ is more favored over $\bar{M}_2$ in explaining the data \cite{Bayes-3}. An empirical scale, commonly referred to as the "Jeffreys scale," is used to evaluate the strength of evidence when comparing two models, $\bar{M_1}$ and $\bar{M_2}$. The Jeffreys scale \cite{Jeffreys-1, Jeffreys-2, Jeffreys-3}
provides a way to assess model preference based on the value of $\left|\ln B\right|$. Specifically, when $\left|\ln R\right|<1$, the evidence is considered inconclusive. Values between $1 \leq\left|\ln R\right|<2.5$ indicate weak evidence, while $2.5 \leq\left|\ln R\right|<5$ suggest moderate evidence. For values $5 \leq\left|\ln R\right|<10$, the evidence is strong, and $\left|\ln R\right| \geq 10$ indicates decisive evidence in favor of one model over the other.
In this paper, the Bayes factors $R$ and the values of $\ln R$ of $\mathrm{TD}$ relative to $\mathrm{R P}_j(j=0,1,2)$ are summarized in Table \ref{table7}.

As shown in Table \ref{table7}, the Bayes factors $R$ of the $\mathrm{TD}$ model relative to $\mathrm{R P}_j(j=0,1,2)$ satisfy $R>1$, indicating that the $\mathrm{TD}$ model is more strongly supported by the microquasar observational data compared to the $\mathrm{R P}_j$ models. Therefore, the $\mathrm{TD}$ model remains the best model. Furthermore, as $\ln R_{\mathrm{TD} / \mathrm{RP}_0}$ and $\ln R_{\mathrm{TD} / \mathrm{RP}_2}$ fall within the range $|\ln R|<1$, which implies that the observational data provide comparable levels of support for the $\mathrm{TD}$ model and the $\mathrm{R P}_0$ and $\mathrm{R P}_2$ models. However, as $\ln R_{\mathrm{TD} / \mathrm{RP}_1}$ with a value exceeding 30, the evidence strongly and decisively favors the $\mathrm{TD}$ model over the $\mathrm{R P}_1$ model. This indicates that the observational data provide almost no support for the $\mathrm{R P}_1$ model.

\begin{table}[ht]
\begin{tabular}{|c|c|c|c|}
\hline & $T D / R P_0$ & $T D / R P_1$ & $T D / R P_2$ \\
\hline$R$ & 1.876 & $2.010 * 10^{14}$ & 1.521 \\
\hline $\ln R$ & 0.629 & 32.934 & 0.419 \\
\hline
\end{tabular}
\caption{The Bayes factors $R$ and  the values of $\ln R$ of $\mathrm{TD}$ relative to $\mathrm{R P}_j(j=0,1,2)$. }  \label{table7}
\end{table}

From Table \ref{table3}, it can be observed that the range of the regular parameter $l^*$, obtained through data fitting, allows the classification of the RSV spacetime into BH (Kerr, regular, and extreme BH) or traversable wormholes. Specifically, when $l^*=l_e^*=\sqrt{-{\alpha^*}^2+2+2 \sqrt{-{\alpha^*}^2+1}}$ (see Section 2 for details), the RSV spacetime represents an extreme BH; when $l^*<l_e^*$, the RSV spacetime corresponds to a black hole; and when $l^*>l_e^*$, the RSV spacetime behaves as a wormhole.
Based on these, we use the error propagation method to calculate the values of $l_e^*$ corresponding to the spin values of three microquasars within the 68\% confidence level, and present the results in Table \ref{table-le}. Additionally, to more intuitively distinguish the astrophysical types of RSV spacetimes under different models, the $l^*$ values within the 68\% confidence level from Table \ref{table3} are also provided in Table \ref{table-le}.

\begin{table}[ht]
\centering
\renewcommand{\arraystretch}{1.5} 
\setlength{\tabcolsep}{5pt} 
\begin{tabular}{|c|c|c|c|c|c|c|c|c|}
\hline
\multicolumn{1}{|c}{} &  & \(ER_0\) & \(ER_1\) & \(ER_2\) & \(RP_0\) & \(RP_2\) & \(TD\) & \(WD\) \\
\hline
\multicolumn{1}{|c}{} & \(l^*\) & \(0.908_{-0.073}^{+0.086}\) & \(3.927 \pm 0.033\) & \(1.850 \pm 0.036\) & \(<0.314\) & \(2.496_{-0.024}^{+0.027}\) & \(4.964 \pm 0.046\) & \(4.358 \pm 0.083\) \\
\hline
\(l_e^*\) & GRO 1655-40 & \(1.397_{-0.039}^{+0.035}\) & \(1.822_{-0.014}^{+0.013}\) & \(1.368_{-0.008}^{+0.008}\) & \(1.959_{-0.000}^{+0.001}\) & \(1.959_{-0.001}^{+0.001}\) & \(1.781_{-0.072}^{+0.069}\) & \(1.896_{-0.022}^{+0.019}\) \\
\cline{2-9}
 & XTE 1550-564 & \(1.336_{-0.122}^{+0.068}\) & \(1.668_{-0.039}^{+0.035}\) & \(1.259_{-0.015}^{+0.015}\) & \(1.934_{-0.008}^{+0.008}\) & \(1.936_{-0.006}^{+0.006}\) & \(1.626_{-0.107}^{+0.069}\) & \(1.777_{-0.056}^{+0.041}\) \\
\cline{2-9}
 & GRS 1915+10 & \(1.493_{-0.131}^{+0.096}\) & \(1.937_{-0.018}^{+0.016}\) & \(1.546_{-0.166}^{+0.012}\) & \(1.987_{-0.008}^{+0.006}\) & \(1.986_{-0.006}^{+0.006}\) & \(1.543_{-0.543}^{+0.191}\) & \(1.964_{-0.024}^{+0.018}\) \\
\hline
\end{tabular}
\caption{Under the model supported by observational data, the summary of results for \(l^*\) and \(l_e^*\).}
\label{table-le}
\end{table}

 As observed in Table \ref{table-le}, for $\mathrm{ER}_0$ and $\mathrm{RP}_0$ models, the observational constraints on RSV regularization parameter are respectively: $l^* = 0.908_{-0.073}^{+0.086}$ and $l^* <0.314$ at $68 \%$ CL, which correspond to the regular or the Kerr BH. For $\mathrm{ER}_1$, $\mathrm{ER}_2$, $\mathrm{RP}_2$, $\mathrm{TD}$, and $\mathrm{WD}$ models, the observational data suggest that RSV objects should be the traversable wormhole, e.g. we have the limits: $l^* =1.850 \pm 0.036$, $l^* =4.964 \pm 0.046$, etc.
 The regularization parameter $l^*$, derived from the fitting of observational data, represents a theoretically viable approach to distinguishing black holes from wormholes. However, it is crucial to acknowledge that this method of distinction depends on the selected HFQPO models, and the selection of different models may influence the fitting outcomes of $l^*$ and its corresponding type of celestial object.

In literature \cite{42}, the authors utilized GRO J1655-40 observational data and the $\chi$-square analysis method, based on the $\mathrm{RP}_0$ model, to constrain the regularization parameter $l^*$ and the spin parameter $\alpha_1$ of the RSV spacetime at $68.3 \% $ $\mathrm{CL}$: $l^*=$ $0.347_{-0.347}^{+0.011}, \alpha_1^*=0.286_{-0.002}^{+0.003}$, which are very close to our fitting results under the same model: $l^*<0.314, \alpha_1^*=0.284_{-0.002}^{+0.001}$. And we also provide the best-fit spin parameter ranges for two other microquasars under the $\mathrm{RP}_0$ model: $\alpha_2^*=$ $0.358 \pm 0.022, \alpha_3^*=0.162 \pm 0.041$. In addition, reference \cite{125} fitted the observational data from NuSTAR spectrum of the Galactic black hole in EXO 1846 031 to limit $l^*$ in the RSV spacetime:  $l^*<0.49$ at $90 \% $ $\mathrm{CL}$. Both constraint results above support RSV as a kerr BH or a regular BH, which is consistent with our fitting results from two HFQPO models ($\mathrm{ER}_0$ and $\mathrm{RP}_0$).
However, in these two models, the spin parameter ranges of three microquasars obtained in this paper (see Table \ref{table3}) are inconsistent with the spin ranges derived from the continuum-fitting method \cite{80,126,127} and the iron-line method  \cite{128,129} (see Table \ref{table8}). This inconsistency could be resolved in the $\mathrm{TD}$ model, at where we have: $\alpha_1^*=0.625_{-0.098}^{+0.081}$, $\alpha_2^*=0.780_{-0.100}^{+0.018}$, $\alpha_3^*=$ $0.840 \pm 0.160$, which can be compared with the results given by the continuum-fitting and the iron-line methods (Table \ref{table8}). It indicates that if the spin values measured by the continuum-fitting and iron-line methods are considered reliable, the RSV spacetime is likely to be a wormhole spacetime.

\begin{table}[ht]
\begin{center}
\begin{tabular}{|c|c|}
\hline Microquasars & Spin \\
\hline GRO 1655-40 $\left(k_1\right)$ & $0.65<\alpha_1^*<0.75^c$ \\
\hline XTE 1550-564 $\left(k_2\right)$ & $0.75<\alpha_2^*<0.77^r$ \\
\hline GRS 1915+105 $\left(k_3\right)$ & $0.97<\alpha_3^*<0.99^{r, c}$ \\
\hline
\end{tabular}
\end{center}
\caption{\label{table8}
     Spin ranges of the three microquasars obtained through the continuum-fitting method \cite{80,126,127} and the iron-line method \cite{128,129}, which are independent of the HFQPO models. The superscripts represent the measurement methods for obtaining the spin range: iron-line ($r$) or continuum fitting ($c$).}
\end{table}
\section{$\text{Conclusion}$}
To solve the problem of singularity of black hole spacetime in gravitational physics, regular black holes have been proposed. Recently, the RSV spacetime promoted by Mazza, Franzin, and Liberati can describe various types of celestial bodies and solve the spacetime singularity problem. When the parameter $l^*=0$ in the line element, the RSV spacetime degenerates to a Kerr black hole. Although Kerr black hole is one of the most popular candidates for astrophysical black holes, many of its properties have yet to be confirmed by observational data. Therefore, studying the theoretical deviations of other different asymptotically stable axisymmetric spacetimes from Kerr black hole is also one of the current research hotspots \cite{62,78,101}. In this paper, we provided the range of circular orbits for particles moving around the epicyclic motion of $(\alpha^* \in[-1,1])$ RSV celestial bodies, tested the stability of these circular orbits, and studied the deviation of the location of the regular black hole, traversable wormhole ISCO, from the Kerr black hole. The study found that for a fixed value of the spin parameter, the position of the ISCO approaches the radial coordinate center as the value of $l^*$ increases, i.e., the traversable wormhole spacetime described by equation (\ref{1}) has an ISCO closer to the central object than Kerr and rotating regular black holes.

Furthermore, given that QPOs can serve as a powerful tool to test gravitational theories, we focused on particles that oscillate in stable circular orbits around central celestial bodies. Consequently, the radial profiles of angular frequency were presented for particles that oscillate around three types of axisymmetric SV celestial bodies with positive rotations, stationary, and negative rotation. We considered several popular HFQPO models, including ER and its variants, RP and its variants, TD, and WD models and studied the possible values of the twin peaks frequency of particles oscillating around traversable wormholes and regular black holes in the RSV spacetime under these models, as well as the deviation from the $3:2$ resonance position for Kerr black holes. The study found that only in the $\mathrm{ER}_0$ and $\mathrm{ER}_5$ models are the positions of resonance for regular black holes and traversable wormholes far away from the radial coordinate center compared to the Kerr black holes. In other nine HFQPO models studied in this paper, the resonance radii of regular black holes and traversable wormholes are all smaller than the resonance radius of Kerr black holes.

Additionally, assuming that these microquasars can be described by the RSV spacetime, we used $\chi^2$ analysis to constrain the regularization parameter $l^*$ and the spin parameters of the three microquasars (GRS 1915+105, XTE 1550-564, and GRO 1655-40) using the observed HFQPO data. Considering that the related result in theory largely depends on the chosen models, it is significant to evaluate various HFQPO models through statistical analysis of observational data. Therefore, we applied the Akaike Information Criterion and the Bayes factor to select models with the help of the observational data. The results showed that $\mathrm{ER}_0$, $\mathrm{ER}_1$, $\mathrm{ER}_2$, $\mathrm{RP}_0$, $\mathrm{RP}_2$,  and $\mathrm{WD}$ models have the same support by observational data ($0 < \Delta_i \leq 2$, $\ln R<1$) as the best model $\mathrm{TD}$. Support for $\mathrm{ER}_3$ and $\mathrm{ER}_5$ decreases notably, and $\mathrm{ER}_4$ and $\mathrm{RP}_1$ receive almost no support from observational data. Concretely, for $\mathrm{ER}_0$ and $\mathrm{RP}_0$ models, the observational constraints on RSV regularization parameter are respectively: $l^* = 0.908_{-0.073}^{+0.086}$ and $l^* <0.314$ at $68 \%$ CL, which correspond to the regular or the Kerr BH. For $\mathrm{ER}_1$, $\mathrm{ER}_2$, $\mathrm{RP}_2$, $\mathrm{TD}$, and $\mathrm{WD}$ models, the observational data suggest that RSV objects should be the traversable wormhole, e.g. we have the limits: $l^* =1.850 \pm 0.036$, $l^* =4.964 \pm 0.046$, etc. The regularization parameter $l^*$, derived from the fitting of observational data, represents a theoretically viable approach to distinguishing black holes from wormholes. And we noticed that the constraint on the spin of microquasars in $\mathrm{TD}$ model in the RSV spacetime could be consistent with the parameters ranges provided by the continuum-fitting and iron line methods. It seems that, if the spin values obtained by the continuum-fitting and iron line methods to be reliable, the RSV spacetime is likely to be a wormhole.

\textbf{\ Acknowledgments }
 The research work is supported by the National Natural Science Foundation of China (12175095,12075109 and 11865012), and supported by  LiaoNing Revitalization Talents Program (XLYC2007047).

We declare: no new data were created or analysed in this study.

\end{document}